\documentclass[twocolumn]{aastex631}

\pdfoutput=1 

\usepackage[utf8]{inputenc}
\usepackage{makecell}
\usepackage{amsmath}
\usepackage{tikz}
\usepackage{hyperref}
\usepackage{soul}
\usepackage{graphicx}
\usepackage{natbib}
\usepackage{afterpage}

\newcommand{\revise}[1]{#1}
\newcommand{\delete}[1]{}
\setstcolor{red}
\newcommand{\fgas}{f_\mathrm{gas}}
\newcommand{\ciii}{{\sc Ciii}]}
\newcommand{\vos}{v_\mathrm{rot}/\sigma_v}
\newcommand{\vosobs}{\Delta v_\mathrm{obs}/\sigma_\mathrm{med}}
\newcommand{\vrot}{v_\mathrm{rot}}
\newcommand{\kms}{\mathrm{km~s^{-1}}}
\newcommand{\msun}{\mathrm{M_\odot}}
\newcommand{\eden}{n_\mathrm{e}}
\newcommand{\rmsub}[2]{#1_\mathrm{#2}}

\received{---}
\revised{---}
\accepted{---}

\shorttitle{
Dynamics of a Galaxy at $z>10$
}
\shortauthors{Xu et al.}

\begin{document}
\title{
Dynamics of a Galaxy at $z>10$ Explored by JWST Integral Field Spectroscopy:\\
Hints of Rotating Disk Suggesting Weak Feedback
%
%
}

\correspondingauthor{Yi Xu}
\email{xuyi@icrr.u-tokyo.ac.jp}

\author[0000-0002-5768-8235]{Yi Xu}
\affiliation{Institute for Cosmic Ray Research, The University of Tokyo, 5-1-5 Kashiwanoha, Kashiwa, Chiba 277-8582, Japan}
\affiliation{Department of Astronomy, Graduate School of Science, the University of Tokyo, 7-3-1 Hongo, Bunkyo, Tokyo 113-0033, Japan}

\author[0000-0002-1049-6658]{Masami Ouchi}
\affiliation{National Astronomical Observatory of Japan, 2-21-1 Osawa, Mitaka, Tokyo 181-8588, Japan}
\affiliation{Institute for Cosmic Ray Research, The University of Tokyo, 5-1-5 Kashiwanoha, Kashiwa, Chiba 277-8582, Japan}
\affiliation{Graduate University for Advanced Studies (SOKENDAI), 2-21-1 Osawa, Mitaka, Tokyo 181-8588, Japan}
\affiliation{Kavli Institute for the Physics and Mathematics of the Universe (Kavli IPMU, WPI), The University of Tokyo, 5-1-5 Kashiwanoha, Kashiwa, Chiba, 277-8583, Japan}

\author[0000-0002-1319-3433]{Hidenobu Yajima}
\affiliation{Center for Computational Sciences, University of Tsukuba, Tennodai, 1-1-1 Tsukuba, Ibaraki 305-8577, Japan}

\author{Hajime Fukushima}
\affiliation{Center for Computational Sciences, University of Tsukuba, Tennodai, 1-1-1 Tsukuba, Ibaraki 305-8577, Japan}

\author[0000-0002-6047-430X]{Yuichi Harikane}
\affiliation{Institute for Cosmic Ray Research, The University of Tokyo, 5-1-5 Kashiwanoha, Kashiwa, Chiba 277-8582, Japan}

\author[0000-0001-7730-8634]{Yuki Isobe}
\affiliation{Waseda Research Institute for Science and Engineering, Faculty of Science and Engineering, Waseda University, 3-4-1, Okubo, Shinjuku, Tokyo 169-8555, Japan}

\author[0000-0003-2965-5070]{Kimihiko Nakajima}
\affiliation{National Astronomical Observatory of Japan, 2-21-1 Osawa, Mitaka, Tokyo 181-8588, Japan}

\author[0009-0000-1999-5472]{Minami Nakane}
\affiliation{Institute for Cosmic Ray Research, The University of Tokyo, 5-1-5 Kashiwanoha, Kashiwa, Chiba 277-8582, Japan}
\affiliation{Department of Physics, Graduate School of Science, The University of Tokyo, 7-3-1 Hongo, Bunkyo, Tokyo 113-0033, Japan}

\author[0000-0001-9011-7605]{Yoshiaki Ono}
\affiliation{Institute for Cosmic Ray Research, The University of Tokyo, 5-1-5 Kashiwanoha, Kashiwa, Chiba 277-8582, Japan}

\author[0009-0008-0167-5129]{Hiroya Umeda}
\affiliation{Institute for Cosmic Ray Research, The University of Tokyo, 5-1-5 Kashiwanoha, Kashiwa, Chiba 277-8582, Japan}
\affiliation{Department of Physics, Graduate School of Science, The University of Tokyo, 7-3-1 Hongo, Bunkyo, Tokyo 113-0033, Japan}

\author{Hiroto Yanagisawa}
\affiliation{Institute for Cosmic Ray Research, The University of Tokyo, 5-1-5 Kashiwanoha, Kashiwa, Chiba 277-8582, Japan}
\affiliation{Department of Physics, Graduate School of Science, The University of Tokyo, 7-3-1 Hongo, Bunkyo, Tokyo 113-0033, Japan}

\author[0000-0003-3817-8739]{Yechi Zhang}
\affiliation{National Astronomical Observatory of Japan, 2-21-1 Osawa, Mitaka, Tokyo 181-8588, Japan}
\affiliation{Institute for Cosmic Ray Research, The University of Tokyo, 5-1-5 Kashiwanoha, Kashiwa, Chiba 277-8582, Japan}
\affiliation{Department of Astronomy, Graduate School of Science, the University of Tokyo, 7-3-1 Hongo, Bunkyo, Tokyo 113-0033, Japan}
\affiliation{Kavli Institute for the Physics and Mathematics of the Universe (Kavli IPMU, WPI), The University of Tokyo, 5-1-5 Kashiwanoha, Kashiwa, Chiba, 277-8583, Japan}

\begin{abstract}
We investigate the dynamics of GN-z11, a luminous galaxy at $z=10.60$, carefully analyzing the public deep integral field spectroscopy (IFS) data taken with JWST NIRSpec IFU. While the observations of the IFS data originally targeted a He{\sc ii} clump near GN-z11, we find that \ciii$\lambda\lambda1907,1909$ emission from ionized gas at GN-z11 is bright and spatially extended significantly beyond the point-spread function (PSF).
The spatially extended \ciii\ emission of GN-z11 shows a velocity gradient, red- and blue-shifted components in the \revise{north and south} directions, respectively, which cannot be explained by the variation of [\ciii$\lambda$1907/\ciii$\lambda$1909 line ratios. \revise{Assuming the velocity gradient is produced by disk rotation,} we perform forward modeling with GalPak$^\mathrm{3D}$, including the effects of PSF smearing and line blending, and obtain a rotation velocity of \revise{$v_\mathrm{rot}=257^{+138}_{-117}$ km s$^{-1}$}, a velocity dispersion of \revise{$\sigma_v=91^{+18}_{-32}$ km s$^{-1}$}, and a ratio of \revise{$v_\mathrm{rot}/\sigma_v=2.83^{+1.82}_{-1.41}$. 
The $v_\mathrm{rot}/\sigma_v$ value would suggest a rotation-dominated disk existing at $z>10$ albeit with the large uncertainties. The rotation velocity agrees with those of numerical simulations predicting a rotating disk formed in the early universe under the condition of mass compaction and weak feedback. While the velocity gradient is consistent with the rotating disk solution, we recognize that galactic outflows can also explain the velocity gradient as well as the extended morphology and the high velocity dispersion found in the outskirt. Higher S/N and resolution data are necessary to conclude the physical origin of the velocity gradient in GN-z11.
%
}

\end{abstract}

\keywords{galaxies: evolution --- galaxies: kinematics and dynamics}

\section{Introduction}

Primordial galaxies evolve into the various types of galaxies we see today involving complex interplay between different processes: accretion of cold gas, minor and major mergers, stellar and active-galactic-nuclei feedback \citep[e.g.,][]{Dekel+09,Hopkins+12,Naab+17,Nelson+19}. On the other hand, many of these baryonic processes operate on physical scales that are well below the resolution of current cosmological simulations, and are mostly calibrated to match observations in the local Universe, leaving potential crisis to explain galaxy evolution in the early universe. Spatially resolved studies of the gas kinematics can be critical to give a consistent picture of how galaxies grow and evolve across cosmic time.

Ground-based surveys using near-infrared integral field unit (IFU) such as SINS \citep{ForsterSchreiber+09,Genzel+11} and KMOS$^\mathrm{3D}$ \citep{Wisnioski+15} provide rich dataset of gas kinematics for star forming galaxies at $z\sim2$. Sub-millimeter observations using ALMA show great potential of investigating gas kinematics at $z>4$. \cite{Neeleman+20,Rizzo+20,Rizzo+21,Tsukui+21} analyze the kinematics of $z\sim4-5$ galaxies and reveal the existence of rotation-dominated disk in the early universe whose rotation velocity $\vrot$ can be as high as ten times of the velocity dispersion $\sigma_v$. \cite{Parlanti+23} further report rotating but turbulent disks at $z>5$ while \cite{Tokuoka+22} identify a potentially rotating disk at $z\sim9$. After the launch of JWST, gas kinematics has been investigated for $z\sim5-8$ galaxies using rest-frame optical emission observed by JWST NIRSpec MSA \citep{deGraaf+23} and NIRCam WFSS \citep{Nelson+23,Li+23}.

Recent JWST observations provide deep spectroscopic data that reveals the properties of $z>10$ galaxies \citep[e.g.,][]{ArrabalHaro+23a,ArrabalHaro+23b,CurtisLake+23,Harikane+24}. GN-z11 is a remarkably luminous high-$z$ galaxies initially identified by \cite{Bouwens+10,Oesch+16}. Recent JWST observations measure a spectroscopic redshift of $10.603$ \citep{Bunker+23}. The NIRCam images reveal the compact size of GN-z11 \citep{Tacchella+23} while the NIRSpec IFU observations identify extended emission \citep{Maiolino+23b,Scholtz+23}. Exploiting the spatial and spectral resolution of 3-dimensional (3d) IFU data, directly analyzing gas kinematics at $z>10$ becomes possible.

In this paper, we analyze the the public data of GN-z11 taken with JWST NIRSpec IFU. While the observations of the IFS data originally targeted a He {\sc ii} clump near GN-z11, we find that \ciii$\lambda\lambda1907,1909$ emission at GN-z11 is bright and spatially extended, which traces the kinematics of ionized gas. This paper is structured as follows. Section \ref{data} explains our observations and dataset. Section \ref{analysis} describes how we analyze the morphology and kinematics of \ciii\ emission. We discuss and summarize our findings in Sections \ref{discuss} and \ref{sum}, respectively. Throughout the paper we adopt a Planck flat $\Lambda$CDM cosmology with $H_0=67.7~\mathrm{km~s^{-1}~Mpc^{-1}}$ and $\Omega_\mathrm{m} = 0.310$ \citep{Planck18}.

\begin{figure*}[htb!]
    \centering
    \includegraphics[width=\linewidth]{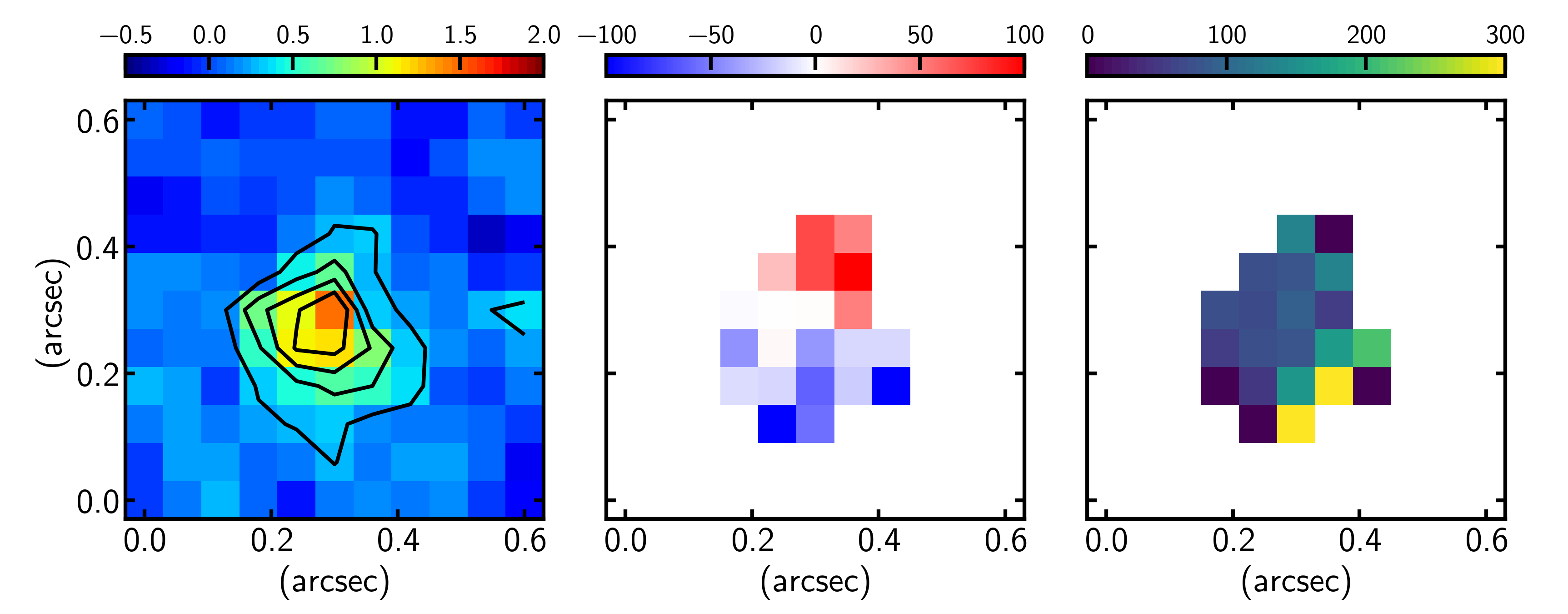}
    \caption{
    Left to right: the moment-0, -1, and -2 maps derived from the \ciii\ emission of GN-z11. 
    \revise{The black contours indicate 2, 4, 6, and 8 $\sigma$ for \ciii\ intensity.}
    The moment-1 and -2 maps are derived by fitting the profile of \ciii~doublets as shown in Figure \ref{fig:fitting}. 
    The colorbar is plotted in an arbitrary unit for the moment-0 map while in the unit of $\mathrm{km~s^{-1}}$ for moment-1 and -2 maps.
    \revise{There exists a velocity gradient from north to south, see text for the details.}
    }
    \label{fig:mom_maps}
\end{figure*}

\section{Observations and data reductions}
\label{data}

In this work we use the data obtained by the NIRSpec IFU \citep[][]{Jakobeson+22,Beoker+22} on JWST. \revise{The data is composed of two parts obtained on 22nd -- 23rd of May 2023 and 17th of April 2024 under the DDT program 4426 (PI: Roberto Maiolino). Details of the first-part observations are presented in \cite{Maiolino+23b} and briefly summarized here}. The observational setup utilised a medium cycling pattern of 10 dithers for a total integration time on source of 10.6 h with the medium-resolution grating/filter pair G235M/F170LP, and 3.3 h with the medium-resolution grating/filter pair G140M/F100LP. These configurations cover the spectral ranges of 0.97--1.89 and 1.66--3.17 $\mu$m, with a nominal spectral resolution of $R\sim1000$. \revise{However, the first-part observations were not centered properly as the guide star selected by the observatory was a binary system, introducing a large offset (1''.4 towards SE; \citealt{Maiolino+23b}). The problem was solved in the second-part observations. The second-part observations utilised the same observational setup with 7 dithers for a total integration time on source of 7.4 h with G235M/F170LP, and 2.3 h with G140M/F100LP.} 

We process the data taken with the grating/filter pair G235M/F170LP using the JWST Science Calibration pipeline \citep{bushouse_2023_10022973} version 1.12.5 under CRDS context \revise{\texttt{jwst\_1256.pmap}}. We make several modifications to the JWST pipeline to increase data quality, which we summarize here. The raw data files are downloaded from the MAST archive (\dataset[10.17909/7wkc-4t02]{10.17909/7wkc-4t02}) and processed with \texttt{calwebb\_detector1}. The resulting count-rate frames are corrected for 1/f noise using \texttt{NSClean} \citep[][]{nsclean}. We make a customized mask to obtain dark areas that are used to fit a background in Fourier space. We then perform the default pipeline steps including wcs-correction, flat-fielding, flux-calibrations, and building 3d data cubes through \texttt{calwebb\_spec2}. \revise{We set the pixels with non-zero DQ flags to \texttt{DO\_NOT\_USE} so they are excluded in the \texttt{cube\_build} step.} We do not apply the \texttt{calwebb\_spec3} process for detecting outliers and stacking frames as we find the data artefacts cannot be perfectly removed (see e.g., \citealt{Perna+23,DEugenio+23}). 
\revise{Instead, we patched the pipeline so that 3d data cubes produced by \texttt{calwebb\_spec2} have the same grid of coordinates, which enables further stacking with median statistics to robustly remove outliers.
We choose a pixel scale of $0''.06$ for the spatial information.
The 3d data cube of each exposure frame is firstly subtracted by background following the method in \cite{Maiolino+23b} before median stacking.}

We carefully inspect each exposure frame to decide whether they should be included in the stacking procedure. \revise{For the firs-part observations that are not centered properly,} we find in 7 frames (\#1, 3, 5, 6, 8, 9, 10) the extended \ciii\ emission with a radius of $\sim0''.3$ is covered within the field-of-view while the other 3 frames (\#2, 4, 7) are not used in this study. 
\revise{Together with the 7 frames from the second-part observations, this study includes 14 frames with a total integration time of 14.8 hours. Note that the wcs of 7 frames from the first-part observations are offsetted which is compensated before stacking. The value of this offset is calculated by making stacked cube for each part and matching the galactic centers.}

In this study, we mainly examine \ciii$\lambda\lambda1907,1909$ that is the emission with highest signal-to-noise ratio (S/N) within the wavelength coverage of G235M/F170LP and G140M/F100LP grating/filter pairs. We make data cube for \ciii\ using the wavelength from 2.2099$\mu$m to 2.2173$\mu$m and the pixels in a $11\times11$ square region centered on the the pixel of \ciii\ intensity peak. The map of the continuum is obtained by sigma-clipping and collapsing continuum channels around \ciii\ emission. We estimate the noise from the root mean squares (rms) of the fluxes in the IFU field of view with no visible sources.

It is critical to characterize the point-spread-function (PSF) and line-spread-function (LSF) of NIRspec IFU in the analysis of kinematics. We utilize the observation data of star G 191--B2B taken on 15th of September 2023 under the calibration program 1537 (PI: Karl Gordon). We reduce the data taken with G235M/F170LP following the procedures mentioned above. The data cube of G 191--B2B is cropped to have the same dimension and wavelength range as that of \ciii\ to represent the PSF. We also apply the same data reduction procedure on the NIRSpec IFU observations of planetary nebula IRAS--05248--7007 conducted on 1st of August 2023 under the calibration program 1492 (PI: Tracy Beck). We investigate the emission line of the planetary nebula in the same manner as \cite{Isobe+23b}. We obtain instrumental broadening of $\rmsub{\sigma}{inst}=8.9~\mathrm{\AA}$ that is consistent with the one of NIRSpec MSA using G235M/F170LP estimated by \cite{Isobe+23b}.

\begin{figure}[htb!]
    \centering
    \includegraphics[width=\linewidth]{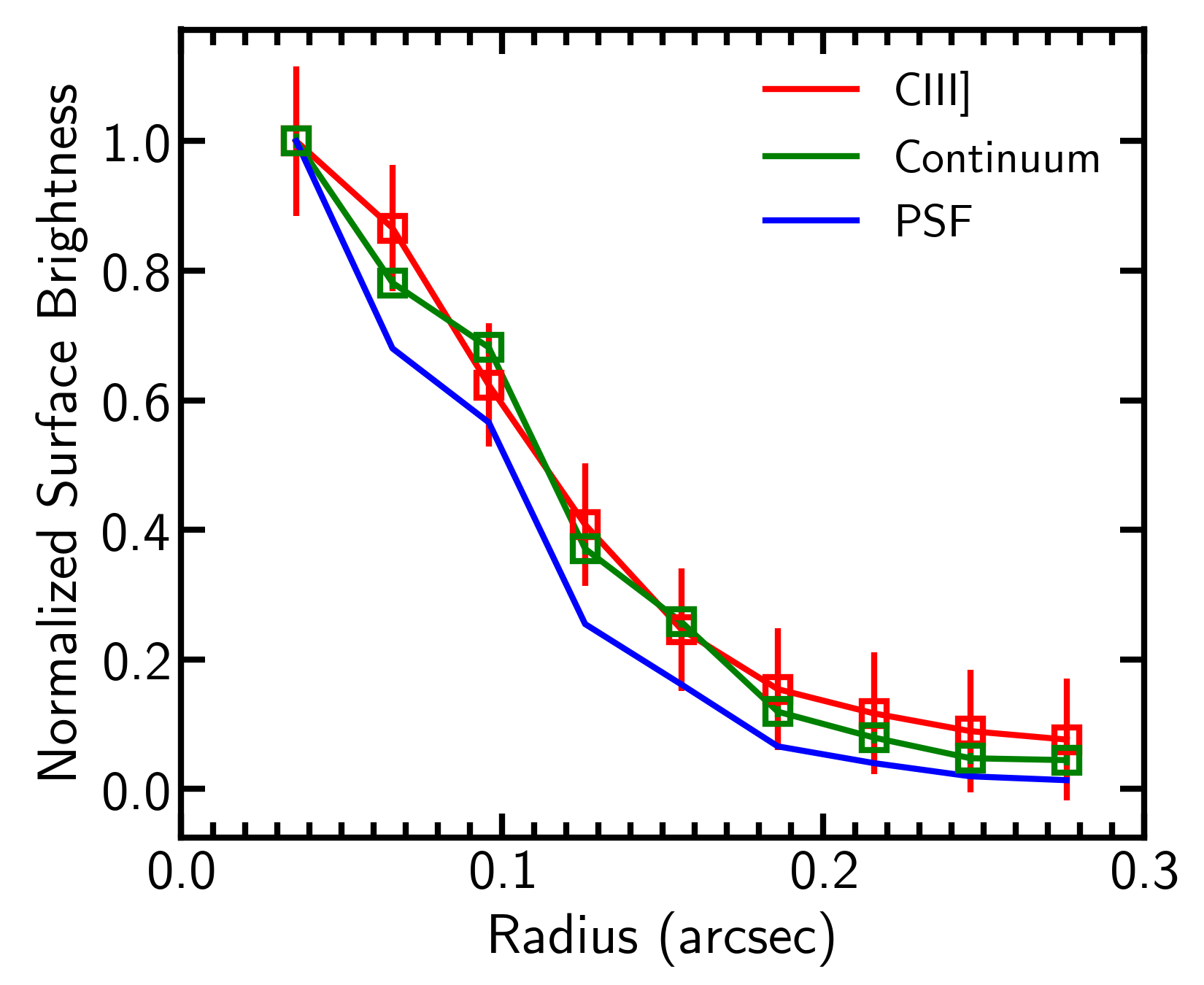}
    \caption{
    Radial profile of \revise{surface brightness} of \ciii\ and continuum compared to the shape of PSF. 
    The \ciii\ profile is more extended than continuum and PSF enabling the analysis of kinematics.
    }
    \label{fig:radial}
\end{figure}

\begin{deluxetable}{lc}[tbh!]
    \label{tab:results}
    \tablecaption{Properties derived from 3D data of \ciii}
    \tablewidth{0pt}
    \tablehead{
      \colhead{Property} & \colhead{Value}
    }
    \startdata
\multicolumn{2}{c}{Morphology Fitting}\\
\hline\\
Half-light radius $R_\mathrm{e}$ [pc] & $221 \pm 42$\\
Position angle $\theta$ [deg] & $112 \pm 28$\\
Axial ratio $b/a$ & $0.60 \pm 0.36$\\
\multicolumn{2}{c}{Observed Kinematic Ratio}\\
\hline\\
Observed velocity difference $\Delta v_\mathrm{obs}$ [km s$^{-1}$] & $347_{-53}^{+63}$\\
Median velocity dispersion $\sigma_\mathrm{med}$ [km s$^{-1}$] & $80_{-10}^{+12}$\\
$\Delta v_\mathrm{obs}/2\sigma_\mathrm{med}$ & $2.16_{-0.37}^{+0.49}$
    \enddata
\end{deluxetable}

\section{Analysis and Results}
\label{analysis}
\subsection{Morphology}
\label{morph}
We make the narrow band image or the so-called moment-0 map of \ciii\ by summing up the data cube in the wavelength direction within the velocity range of $-500~\kms$ to $800 ~\kms$ around the centroid of [\ciii$\lambda1907$. 
\revise{In the left panel of Figure \ref{fig:mom_maps} we show the moment-0 map together with the 2, 4, 6, and 8 $\sigma$ contours. Similar to the one presented in \cite{Maiolino+23b}, the 4 and 6 $\sigma$ contours are spatially extended from northwest to southeast, which is consistent with the position angle of NIRCam images modelled by \cite{Tacchella+23}. On the other hand, we find extended component in north and south direction from the 2 $\sigma$ contour. With the current data, we cannot conclude whether the outskirt of \ciii\ morphology is given by an extended disk or a separated component such as galactic outflows.}
We fit the moment-0 map with a Sersic model using  \texttt{galfit} \citep{Peng+02,Peng+10}. The shape of PSF we measure in Section \ref{data} is convolved with the model during the fitting. We obtain a Sersic index of $n<0.5$ but with large uncertainties. We thus adopt the fiducial fitting result with a $n=1$ Sersic profile and obtain a half-light radius of \revise{$R_\mathrm{e}=221 \pm 42~\mathrm{pc}$} as shown in Table \ref{tab:results}. \cite{Tacchella+23} find the NIRCam images of GN-z11 are composed of a central unresolved component and a extended exponential component. Effective radius of \ciii~ is \revise{consistent with} the extended component ($R_\mathrm{e}=196\pm12~\mathrm{pc}$) that is composed of more matured stellar population than the central region. In Figure \ref{fig:radial}, we show the radial profile of \ciii\ \revise{surface brightness that is more extended than that of PSF}. The surface brightness profile is derived by integrating the fluxes within a series of annulus apertures that oversample the moment-0 map. From the surface brightness profile, we derive the half-light radius of $R_\mathrm{e}\sim300$ pc after subtracting the size of PSF, which is broadly consistent with the once derived from galfit fitting. The \ciii\ morphology is clearly extended and resolved, which enables further kinematic analysis.

\begin{figure}
    \centering
    \includegraphics[width=\linewidth]{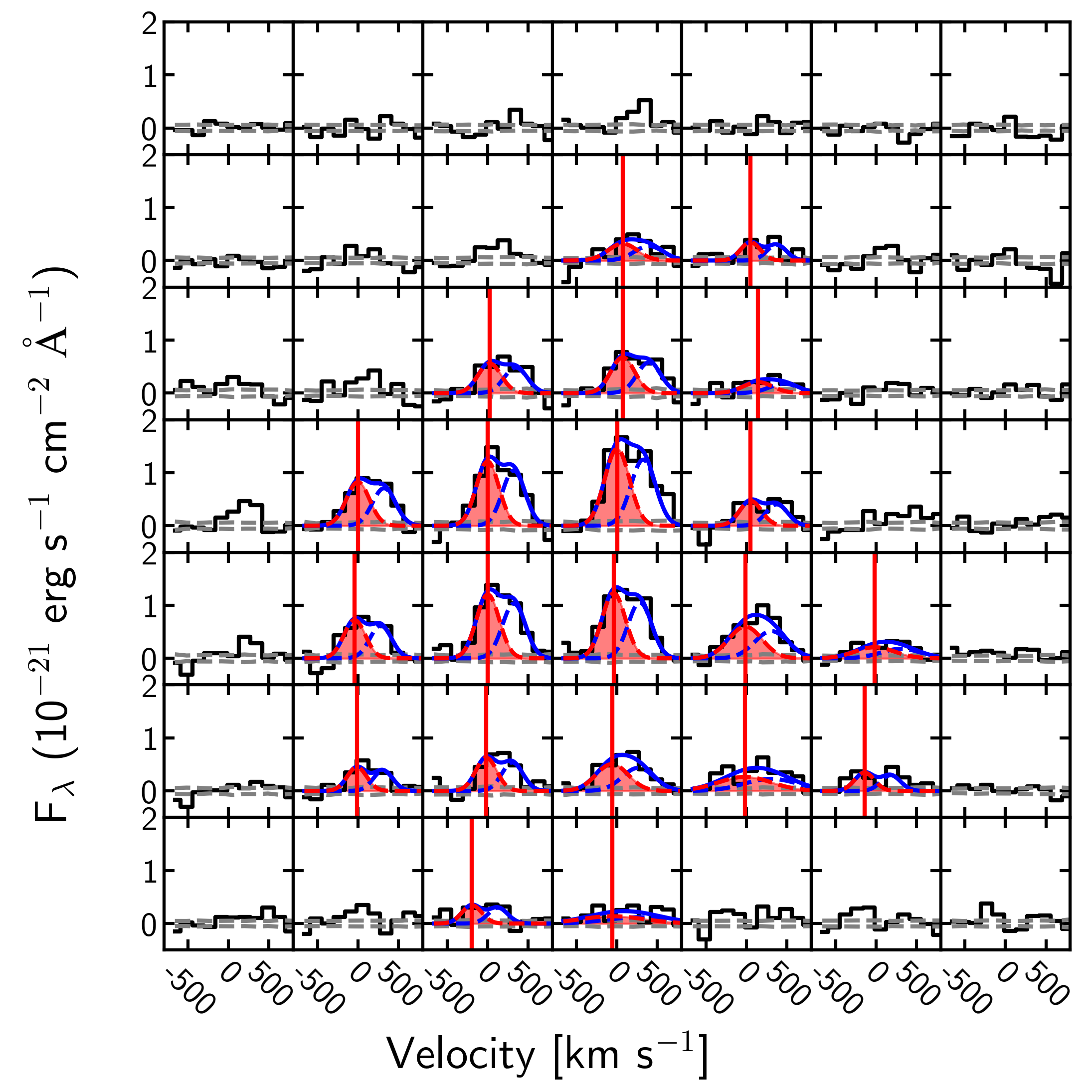}
    \caption{Spectra of \ciii\ in the central $49=7\times7$ pixels of Figure \ref{fig:mom_maps}. In each pixel with sufficient S/N, we overplot the best-fit profile (blue solid curve) composed of two Gaussian components (dashed curves). \revise{Red shades highlight the [{\sc Ciii}]$\lambda1907$ component that is used to derive the line-of-sight velocity and velocity dispersion with the red vertical line indicating the central velocity.}}
    \label{fig:fitting}
\end{figure}

\begin{figure*}[htb!]
    \centering
    \includegraphics[width=\linewidth]{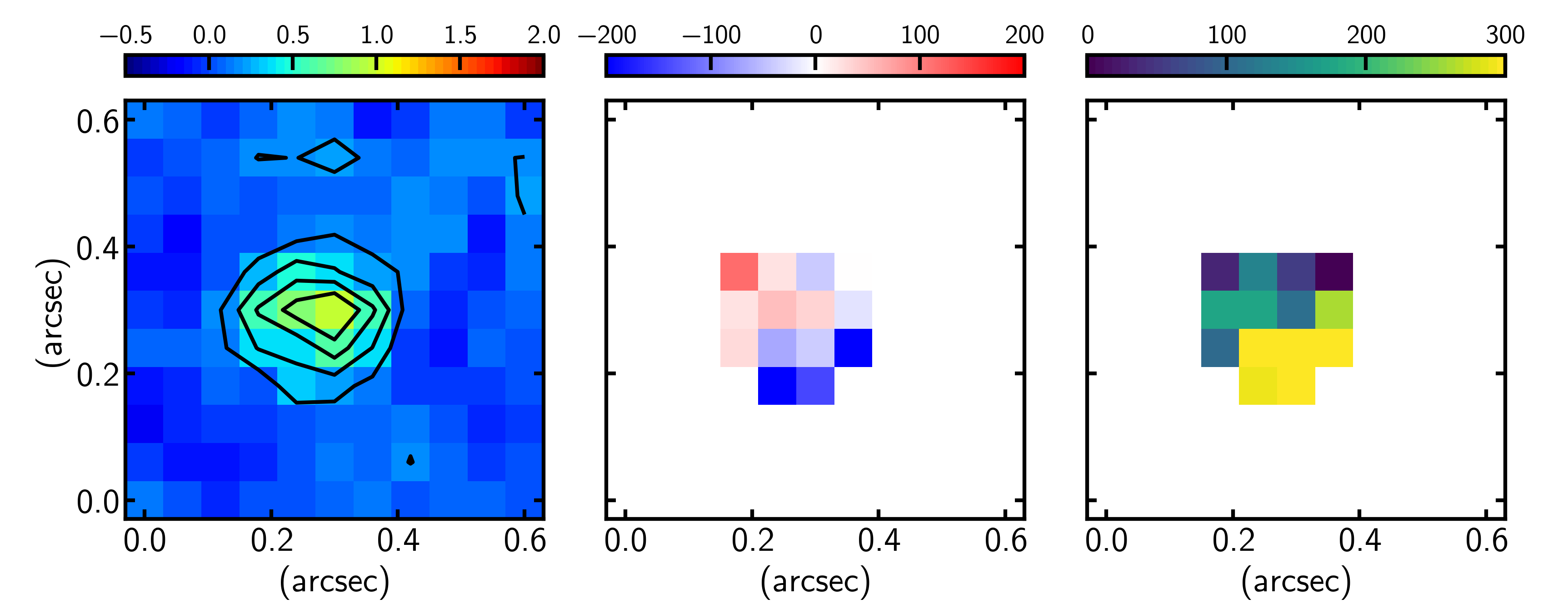}
    \caption{
    Same as Figure \ref{fig:mom_maps} but for {\sc N iv]}$\lambda1486$. 
    }
    \label{fig:NIV_mom_maps}
\end{figure*}

\begin{figure*}[htb!]
    \centering
    \includegraphics[width=\linewidth]{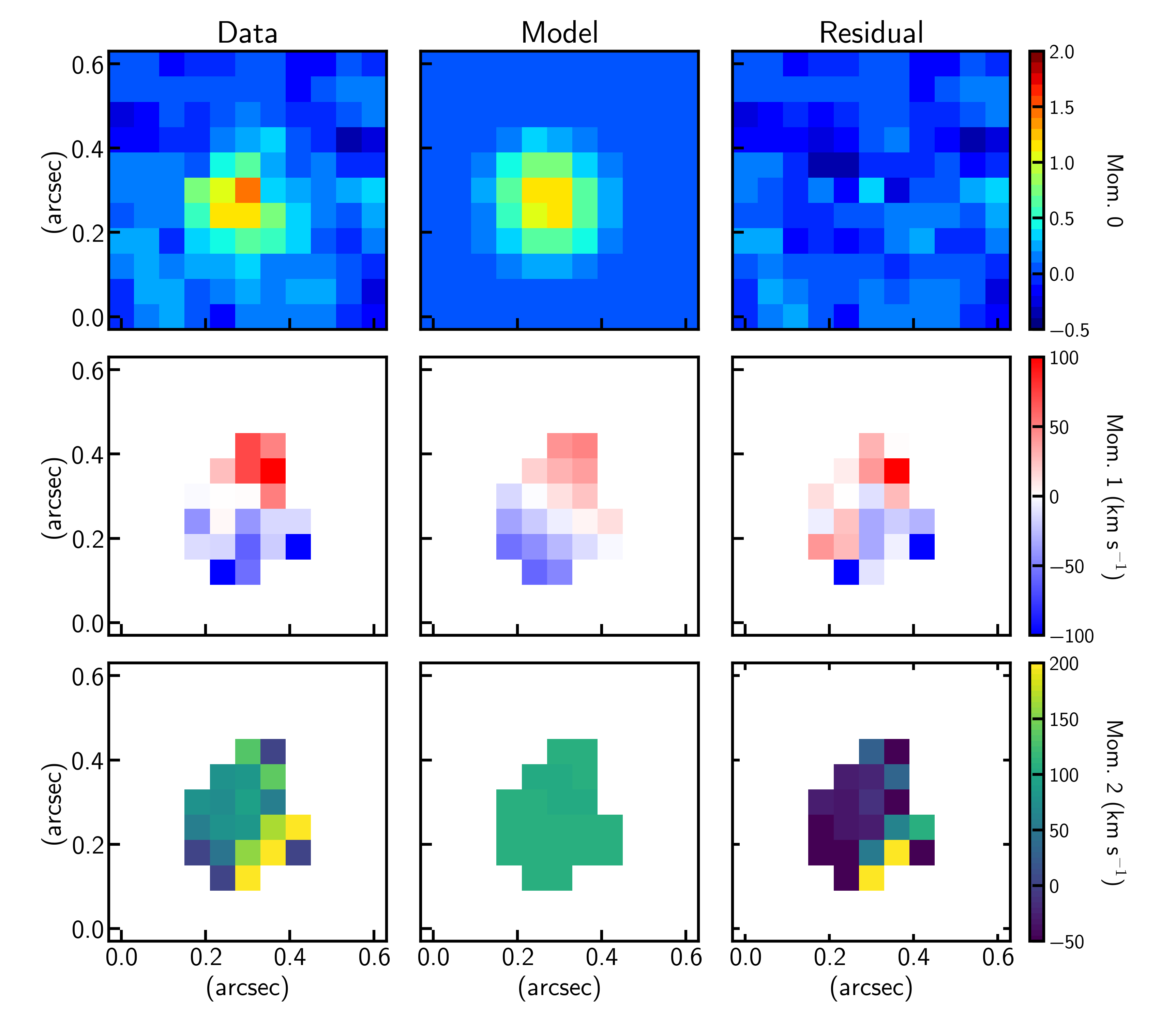}
    \caption{
    Moment maps compared with the best-fit rotation disk model. Left to right: observed data, GalPak$^\mathrm{3D}$ model, and the residuals.  
    }
    \label{fig:mom_maps_glpk}
\end{figure*}

\begin{figure*}[htb!]
    \centering
    \includegraphics[width=\linewidth]{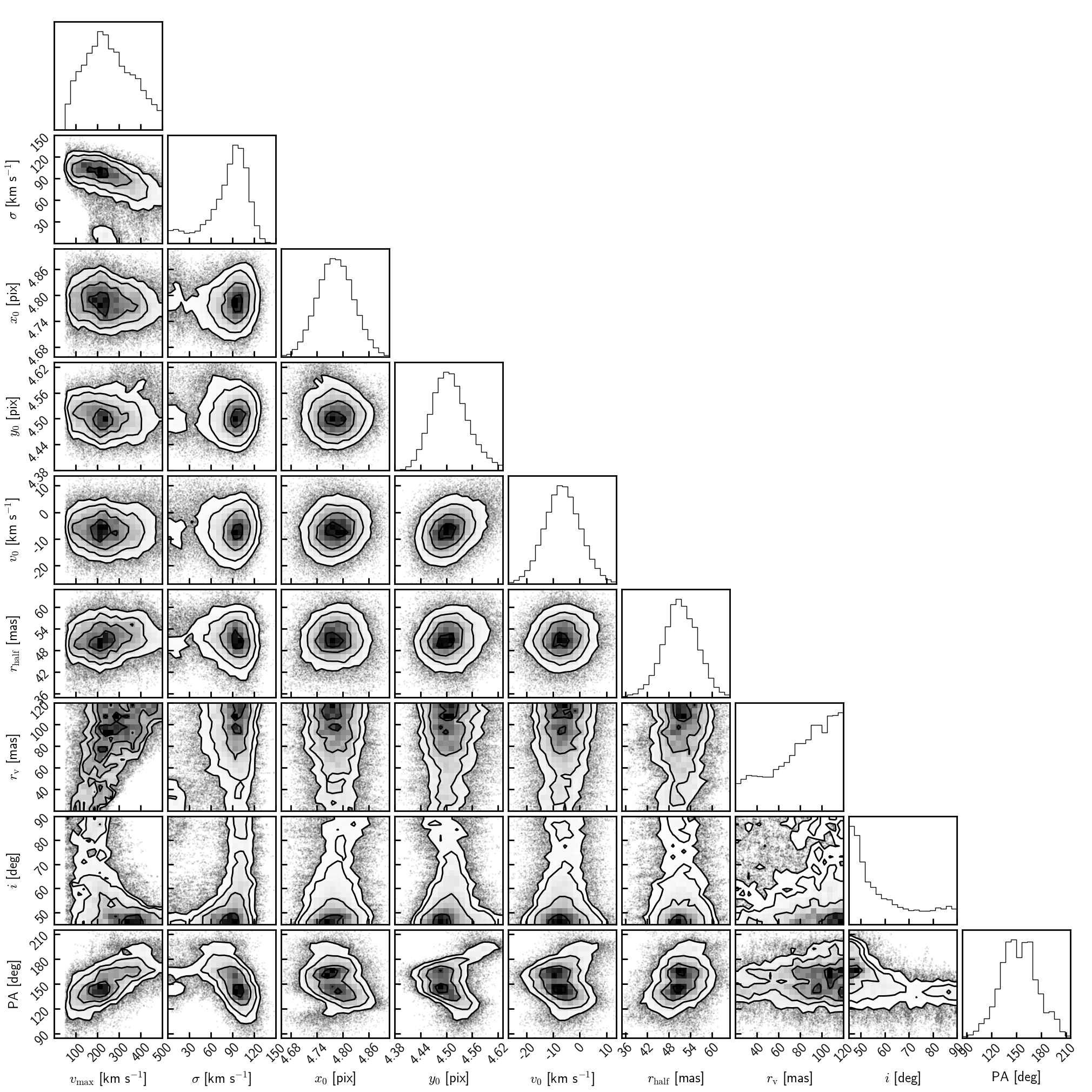}
    \caption{
    \revise{Corner plot for the posterior distribution of parameters of the rotating disk model.}
    }
    \label{fig:corner_glpk}
\end{figure*}

\subsection{Kinematics}
\label{kinematics}
We examine the \ciii\ kinematics by calculating the moment-1 and -2 maps that represent the line-of-sight velocity and velocity dispersion, respectively. Instead of the method by summing up the velocity weighted by the fluxes \citep[e.g.,][]{Walter+08}, we adopt a method of profile fitting to the spectrum in each pixel as shown in Figure \ref{fig:fitting}. The profile fitting method is more robust against the fluctuations of flux especially at the wavelength channels that correspond to large velocity offset. We select spaxels with \revise{S/N$>2$} that can be fitted reliably. Because \ciii\ emission is a doublet, we fit the emission line with two Gaussian components with the constraints of: 1) flux ratio of \revise{[{\sc Ciii}]$\lambda1907/${\sc Ciii}]$\lambda1909=1.16$} (\citealt{Kewley+19}; see Section \ref{ne_discuss} for further discussions), 2) velocity offset between the two centroids fixed to the one at rest-frame, 3) identical line widths larger than the LSF dispersion of $\rmsub{\sigma}{inst}=8.9~\mathrm{\AA}$. By conducting the profile fitting to the integrated spectrum of S/N$>2$ spaxels, we find the central wavelength of [\ciii$\lambda1907$ to be \revise{$2.2124\mu$m} corresponding to the redshift of \revise{$z=10.603$} that is defined as the systematic velocity. As shown in the middle panel of Figure \ref{fig:mom_maps}, there is a clear velocity gradient from \revise{north to south}. We measure a maximum observed velocity difference of \revise{$\Delta v_\mathrm{obs}=347_{-53}^{+63}~\kms$} and a median velocity dispersion of \revise{$\sigma_\mathrm{med}=80_{-10}^{+12}~\kms$}. The velocity dispersion is the line width $\sigma$ quadratically subtracted by the $\sigma_\mathrm{inst}$. This value is likely an upper limit because the emission line only marginally resolved compared to the LSF dispersion of $\sim120~\kms$. \revise{The uncertainties are measured in a Monte-Carlo manner by creating 1000 random data cubes.} Here we report the observed kinematic ratio of \revise{$\Delta v_\mathrm{obs}/2\sigma_\mathrm{med}=2.16_{-0.37}^{+0.49}$ which can be possibly explained by a rotation dominated disk. We recognise that galactic outflows can also lead to the observed velocity gradient.  We note that the direction of velocity gradient (PA$\sim160^\circ$) is different from the extended disk identified from NIRCam images (PA$=34^\circ$, \citealt{Tacchella+23}) but roughly consistent with the polar direction, although the extended disk is only marginally detected. }

\revise{We have conducted similar analysis to other emission lines detected with the G235M grating IFU data. We find the integrated S/N for {\sc N iv]}$\lambda1486$, {\sc N iii]}, and He {\sc ii}$\lambda1640$ are 6, 7, and 2, respectively, all of which are lower than S/N $=11$ for \ciii. S/N of He {\sc ii} is too low to extract useful information. {\sc N iii]} is a multiplet primarily composed of four semi-forbidden lines, which does not allow reliable discussions on kinematics.  Although {\sc N iv} is usually contributed by the forbidden {\sc [N iv]}$\lambda1483$ line, the forbidden line is negligibly weak due to high density \citep[see also][]{Maiolino+24}. The single {\sc N iv]}$\lambda1486$ line may be used to trace kinematics while S/N is lower than \ciii\ doublet. As shown in Figure \ref{fig:NIV_mom_maps}, we find similar moment-1 and -2 maps to \ciii\ with red- and blue-shifted velocities on north and south, respectively.}

Assuming the velocity gradient seen in the \ciii\ moment-1 map is given by disk rotation, we conduct forward modelling using a software named GalPak$^\mathrm{3D}$ \citep[][]{Bouche+15}. GalPak$^\mathrm{3D}$ allows us to fit 3d data cube of doublets assuming a fixed flux ratio and convolve the dynamical model with the PSF shape we derive in Section \ref{data}. \revise{For the fiducial fit,} we choose a thick disk model with the disk height equals to $30\%$ of the effective radius. The surface brightness is an exponential function of radius, i.e., Sersic profile with a Sersic index of $n=1$. The choice of the surface brightness profile and disk height is consistent with the modelling result of \ciii\ morphology. We choose the arctan rotation curve with two parameters, maximum rotation velocity ($\vrot$) and turn-over radius ($r_\mathrm{v}$):
\begin{equation}
v(r) = \vrot \frac{2}{\pi}\arctan(r/\rmsub{r}{v}).
\label{eq:rotation_curve}
\end{equation}
The disk model consists of 10 parameters in total that are x- and y-coordinates of the center, total flux, effective radius ($R_\mathrm{e}$), $r_\mathrm{v}$, inclination ($i$), position angle, systemic velocity, $\vrot$, and intrinsic dispersion ($\sigma_v$). \revise{The fitting is performed with Markov chain Monte Carlo (MCMC).} We fix the prior to the uniform distributions whose lower and upper boundaries are given by the galfit modeling of \ciii\ morphology. The intrinsic dispersion is free from the broadening given by instrument, the local isotropic velocity dispersion driven by disk self-gravity, the mixture of the line-of-sight velocities due to the disk thickness \cite[][]{Bouche+15}, and the blending of \ciii~doublets. The best-fit model and the residuals of moment maps are presented in Figure \ref{fig:mom_maps_glpk}. While the velocity gradient can be explained by the best-fit model, there still remain residuals on the moment-1 maps possible given by complex dynamical component that is not resolved by the current data. \revise{The dynamical properties are shown in Column (1) of Table \ref{tab:models} with the $1\sigma$ uncertainties obtained from the MCMC chain.}
We obtain \revise{$\vos=2.83^{+1.82}_{-1.41}$ larger than unity suggestive of} a rotation dominated disk in GN-z11. 

\revise{Figure \ref{fig:corner_glpk} shows the posterior distribution of the fitting parameters. While most parameters are well constrained with reasonable dispersions, inclination $i$ and turn-over radius $\rmsub{r}{v}$ are not well constrained. We constrain inclination to be larger than $45^\circ$. If we allow $i$ to vary from 0 to $90^\circ$, the posterior distribution of $i$ shows a bimodality with maxima at $i\sim30^\circ$ and $80^\circ$ corresponding to $\vrot\sim800~\kms$ and $200~\kms$, respectively. We thus conduct fitting with relatively large $i$ which gives more conservative rotation velocity and has better agreement with the morphology. Turn-over radius determines the radius where the rotation curve becomes flat. A large $\rmsub{r}{v}$ would indicate that the rotation curve continues to rise beyond the observed region leading to large $\vrot$. As shown in Figure \ref{fig:corner_glpk}, $\vrot$ mildly increases with larger $\rmsub{r}{v}$. Since we cannot probe the flat part of the rotation curve with the current data, we fit $\rmsub{r}{v}$ within the range of $0.5\rmsub{R}{e}$ to $2\rmsub{R}{e}$.}

\revise{We further examine how $\vrot$ and $\sigma_v$ would vary with models of different assumptions. Starting from the fiducial model, we test different models changing 1) shape of PSF, 2) disk thickness, 3) rotation curve formula. The results of different models are presented in Table \ref{tab:models}. Since the observations of the star we choose as PSF may not have identical conditions to the observations of GN-z11, we use \texttt{WebbPSF} \citep{WebbPSF} to simulate the point spread function specifying the observational setup. We note that the size of PSF given by \texttt{WebbPSF} may be underestimated, which is known for the case of NIRCam \citep[e.g.,][]{Zhuang+24}. As one may expect, we obtain larger $\rmsub{R}{e}$ and $\vrot$ since PSF given by \texttt{WebbPSF} is narrower than the one of fiducial model. For the disk thickness, we perform forward modelling for a thin disk whose disk height is equals to $10\%\rmsub{R}{e}$ because GalPak$^\mathrm{3D}$ does not implement an infinitely thin disk. For the rotation curve, the arctan function chosen for the fiducial model approaches a flat maximum velocity towards the outer region where the contribution of dark matter becomes significant \citep[e.g.,][]{deBlok+08}. As a comparison we test the rotation curve of Freeman disk given by an exponential baryonic disk \citep{Freeman70}:
\begin{equation}
v^2(y=r/\rmsub{R}{d}) = G\frac{\rmsub{M}{d}}{\rmsub{R}{d}} y^2 [I_0(y)K_0(y) - I_1(y)K_1(y)],
\end{equation}
where $\rmsub{R}{d}$ is the disk scale length, $\rmsub{M}{d}$ is the disk mass, and $I_n$ and $K_n$ are modified Bessel functions of the first and second kind, respectively.
The rotational velocity for Freeman disk is defined as the peak velocity at $y=1.1$.
Because the fiducial model suggests the flat part is possibly not probed by the current data, $\vrot$ given by the fiducial models is larger than that of Freeman disk whose velocity peaks in a smaller radius.
While the current data does not allow us to distinguish between different models suggested by the similar reduced chi squares, the values of $\vrot$ and $\sigma_v$ are in general within the uncertainty of the fiducial model.}

\delete{The results above are based on the stacked data cube for 4 healthy frames. Although we find potential data artefacts in the other 3 frames (see Section }\ref{data}\delete{), we also conduct the same analysis on the seven-frame stack and obtain similar results, such as $\Delta\rmsub{v}{obs}=238_{-67}^{+78}~\kms$, $\rmsub{\sigma}{med}=77_{-77}^{+68}~\kms$, $\vrot=361_{-179}^{+184}~\kms$, and $\sigma_v=75_{-52}^{+22}~\kms$. We adopt the results derived from the four-frame stack as the fiducial values for further discussions.}

\begin{deluxetable*}{lcccc}[tbh!]
    \label{tab:models}
    \tablecaption{Forward modelling results assuming different models}
    \tablewidth{0pt}
    \tablehead{
      \colhead{Property} & \colhead{Fiducial} & \colhead{WebbPSF} & \colhead{Freeman disk} & \colhead{Thin disk}\\
      \colhead{(1)} & \colhead{(2)} & \colhead{(3)} & \colhead{(4)} & \colhead{(5)}
    }
    \startdata
Half-light radius $R_\mathrm{e}$ [pc] & $209^{+19}_{-18}$ & $368^{+18}_{-15}$ & $199^{+21}_{-18}$ & $143^{+17}_{-24}$ \\
Position angle $\theta$ [deg] & $152 \pm 21$ & $170^{+7}_{-9}$ & $155^{+20}_{-22}$ & $180 \pm 0$ \\
Inclination $i$ [deg] & $54^{+18}_{-7}$ & $27^{+4}_{-6}$ & $33^{+22}_{-14}$ & $54 \pm 0$ \\
Rotational velocity $v_\mathrm{rot}$ [km s$^{-1}$] & $257^{+138}_{-117}$ & $325^{+124}_{-115}$ & $141^{+64}_{-53}$ & $319^{+94}_{-69}$ \\
Velocity dispersion $\sigma_v$ [km s$^{-1}$] & $91^{+18}_{-32}$ & $88^{+16}_{-23}$ & $87^{+17}_{-27}$ & $88 \pm 13$ \\
$v_\mathrm{rot}/\sigma_v$ & $2.83^{+1.82}_{-1.41}$ & $3.67^{+1.69}_{-1.46}$ & $1.62^{+0.89}_{-0.70}$ & $3.64^{+1.20}_{-0.95}$ \\
Reduced chi-square $\chi^2_\nu$ & 2.31 & 2.15 & 2.31 & 2.23 \\
    \enddata
\end{deluxetable*}

\begin{figure}[htb!]
    \centering
    \includegraphics[width=\linewidth]{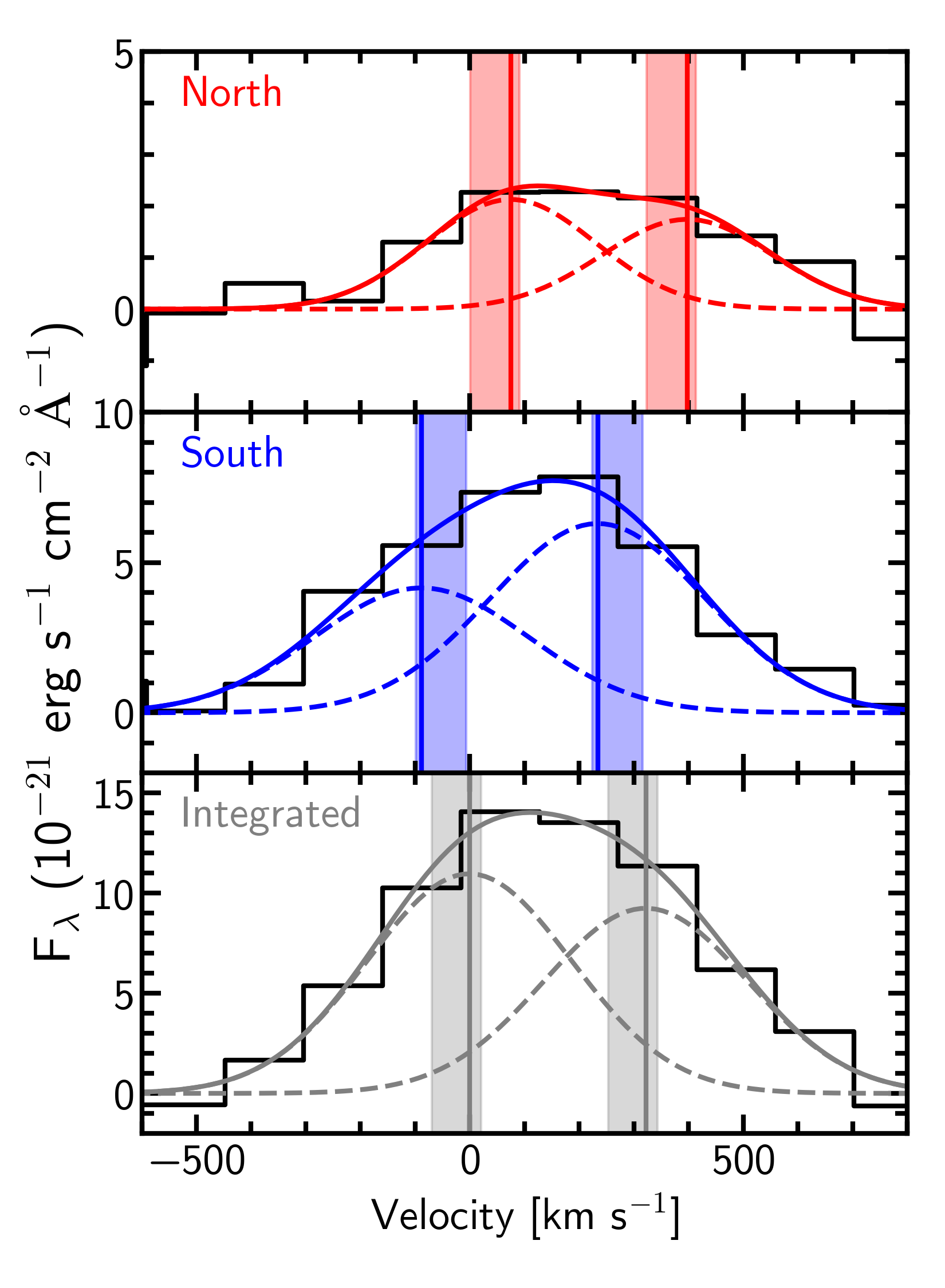}
    \caption{
        Integrated \ciii\ spectra of S/N$>2$ pixels on the \revise{northern (top), southern (middle), and the entire area (bottom)} of GN-z11. In the same manner as Figure \ref{fig:fitting}, we show the best-fit profile but with flux ratio [\ciii$\lambda$1907/\ciii$\lambda$1909 as a free parameter. The solid vertical lines indicate the centroids of [\ciii$\lambda1907$ and \ciii$\lambda1909$ obtained from the best-fit profile. The left and right boundaries of the shaded areas are obtained from the best-fit profiles with flux ratio fixed to $0.588$ ($\rmsub{n}{e}=10^5~\mathrm{cm^{-3}}$) and $1.48$ ($\rmsub{n}{e}=100~\mathrm{cm^{-3}}$), respectively. The velocity offset between the \revise{northern and southern} pixels cannot be explained by the variation of flux ratio \revise{unless $\eden$ changes by three orders of magnitudes}.
    }
    \label{fig:flux_ratio}
\end{figure}

\subsection{Velocity gradient explained by flux ratio variation?}
\label{ne_discuss}


The flux ratio of [\ciii$\lambda$1907/\ciii$\lambda$1909 is sensitive to the electron density between $\rmsub{n}{e}=10^2-10^5~\mathrm{cm^{-3}}$. We examine the flux ratio based on the ionization models of \citep{Kewley+19} with electron temperature $\rmsub{T}{e}=10^4~\mathrm{K}$. In Figure \ref{fig:flux_ratio}, we presents spectra of \ciii\ integrated for all S/N$>2$ pixels, those on the \revise{north}, and those on the \revise{south} together with the best-fit profiles with flux ratio [\ciii$\lambda$1907/\ciii$\lambda$1909 as a free parameter. To examine the influence of flux ratio on the central wavelengths, we also conduct fitting with flux ratio [\ciii$\lambda$1907/\ciii$\lambda$1909 fixed to $0.588$ ($\rmsub{n}{e}=10^5~\mathrm{cm^{-3}}$) and $1.48$ ($\rmsub{n}{e}<100~\mathrm{cm^{-3}}$). Changing the flux ratio results in a velocity difference of $\lesssim100~\kms$. The velocity difference between the \revise{northern to southern pixels can only be explained by electron density variation from $<100~\mathrm{cm^{-3}}$ to $10^5~\mathrm{cm^{-3}}$ as shown in Figure \ref{fig:flux_ratio}, which is unlikely.}

We obtain [\ciii$\lambda1907$/\ciii$\lambda1909=1.16$ corresponding to $\eden\sim10^4~\mathrm{cm^{-3}}$ from the best-fit profile of the integrated spectrum. Relatively high electron density of $\gtrsim10^3~\mathrm{cm^{-3}}$ is consistent with reshift evolution of $\eden$ up to $z\sim9$ discussed in \cite{Isobe+23b}. However, we obtain large $1\sigma$ uncertainty $=0.54$ on the flux ratio based on the integrated spectrum. The models with $\rmsub{n}{e}=10^5~\mathrm{cm^{-3}}$ or $\rmsub{n}{e}=100~\mathrm{cm^{-3}}$ cannot be ruled out with $\chi^2$ statistics or Bayesian evidence. We choose a flux ratio of \revise{1.16 corresponding to $\eden=10^4~\mathrm{cm^{-3}}$} to measure the velocities and conduct dynamical modelling.

\begin{figure*}[htb!]
    \centering
    \includegraphics[width=0.75\linewidth]{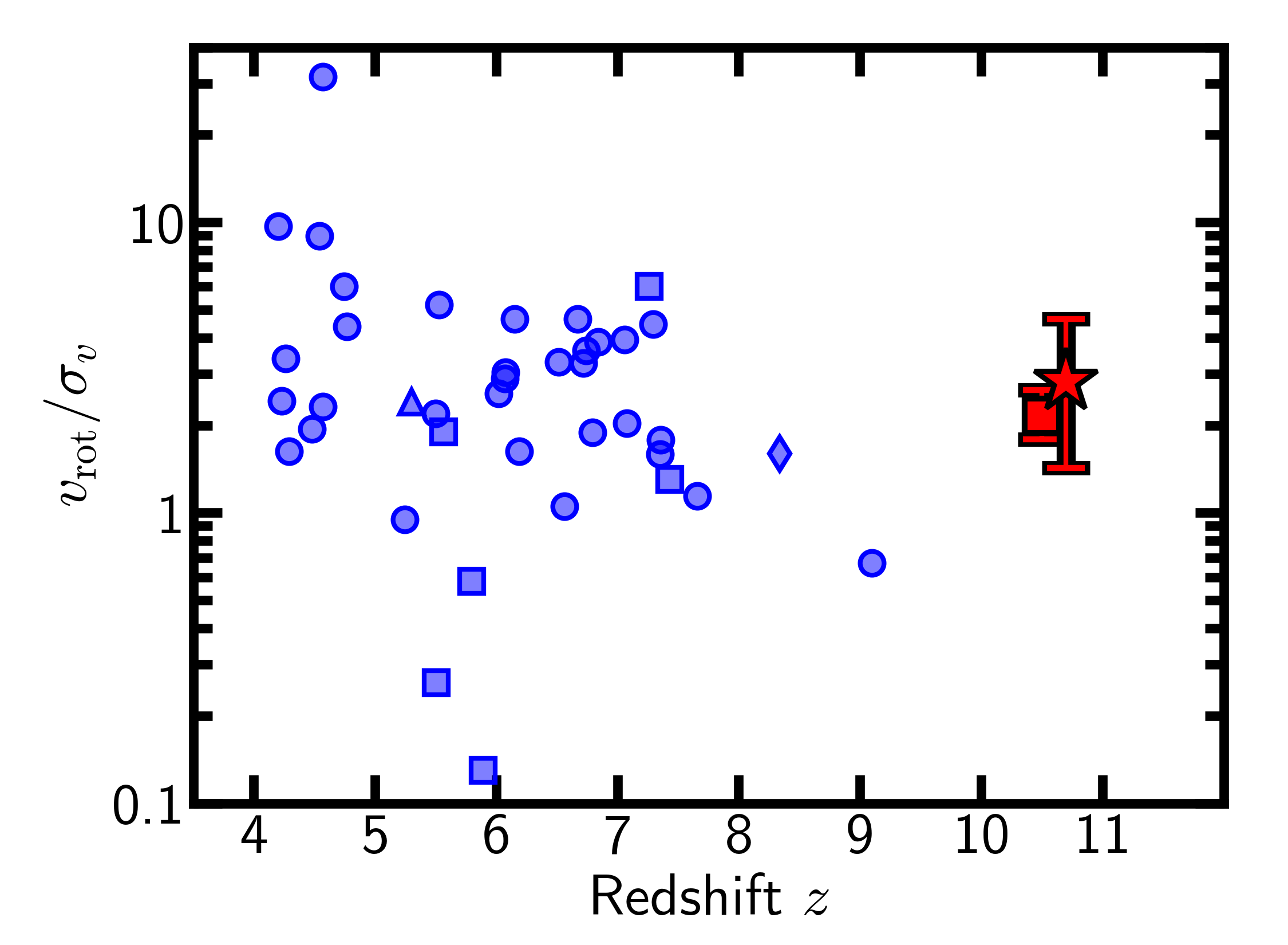}
    \caption{
    $\vos$ of GN-z11 compared to those obtained by previous studies at different redshifts. Our results of GN-z11 are shown with the red star and square that are $\vos$ obtained with GalPak$^\mathrm{3D}$ modelling and the observed kinematic ratio $\vosobs$, respectively. The blue circles are galaxies whose [{\sc Cii}] kinematics is measured with ALMA taken from \cite{Neeleman+20,Rizzo+20,Rizzo+21,Tsukui+21,Fraternali+21,Parlanti+23}. The blue squares are recent JWST results using NIRSpec MSA taken from \cite{deGraaf+23}, while the blue triangles and diamonds are taken from \cite{Nelson+23} and \cite{Li+23}, respectively, who use NIRCam WFSS.
    }
    \label{fig:voversigma_z}
\end{figure*}

\section{Discussions}
\label{discuss}

\subsection{Rotation disk at $z>10$}


The ratio between ordered rotation and turbulent motion is often used to indicate a rotation disk. Our estimations of $\vos$ are displayed in Figure \ref{fig:voversigma_z} with a comparison to previous results at different redshifts. Both the observed kinematic ratio and the intrinsic $\vos$ agree with the scenario of a rotation dominated disk. Our results are comparable to previous rotating galaxies identified by ground-based sub-millimeter observations \citep{Rizzo+20,Rizzo+21,Tsukui+21,Herrera-Camus+22,Tokuoka+22,Parlanti+23}, rest-frame optical spectroscopy using JWST NIRSpec MSA \citep{deGraaf+23}, and NIRCam WFSS \citep{Nelson+23,Li+23}. Nevertheless, it is surprising to find a rotation disk at $z>10$ because galaxies are more likely to be turbulent with increasing redshift due to feedback from massive stars and supernovae \citep[e.g.,][]{Pillepich+19,Yajima+22,Yajima+23}. Simulations of \cite{Yajima+17} show that galactic disks are destroyed due to supernova feedback, while galaxies in simulations with no feedback or lower SF efficiency models can sustain a galactic disk. Recent JWST results of massive luminous galaxies found at $z>10$ \citep[e.g.,][]{Donnan+23a,Harikane+23a,Harikane+24} suggest lack of suppression on star formation. High electron density such found by \cite{Isobe+23b} could be an evidence of efficient cooling in which case feedback becomes inefficient. \cite{Dekel+23} and \cite{ZhaozhouLi+23} also discuss that high densities and low metallicities in galaxies at $z\gtrsim10$ result in a high star formation efficiency with feedback-free starbursts \revise{(hereafter FFB), which is discussed in details below. \cite{Fujimoto+24} identify a rotating system with numerous clumps that may form under the weak feedback scenario. As for AGN feedback, although GN-z11 shows signatures of AGN-driven outflows (e.g., {\sc C iv} absorption, \citealt{Maiolino+24}), feedback from the AGN hosted by GN-z11 as well as many other high-z AGNs is less efficient compared to local ones \citep{Maiolino+24b}.} The scenario of weak feedback is consistent with GN-z11 being luminous and rotational dominated.


\begin{figure*}[htb!]
    \centering
    \includegraphics[width=\linewidth]{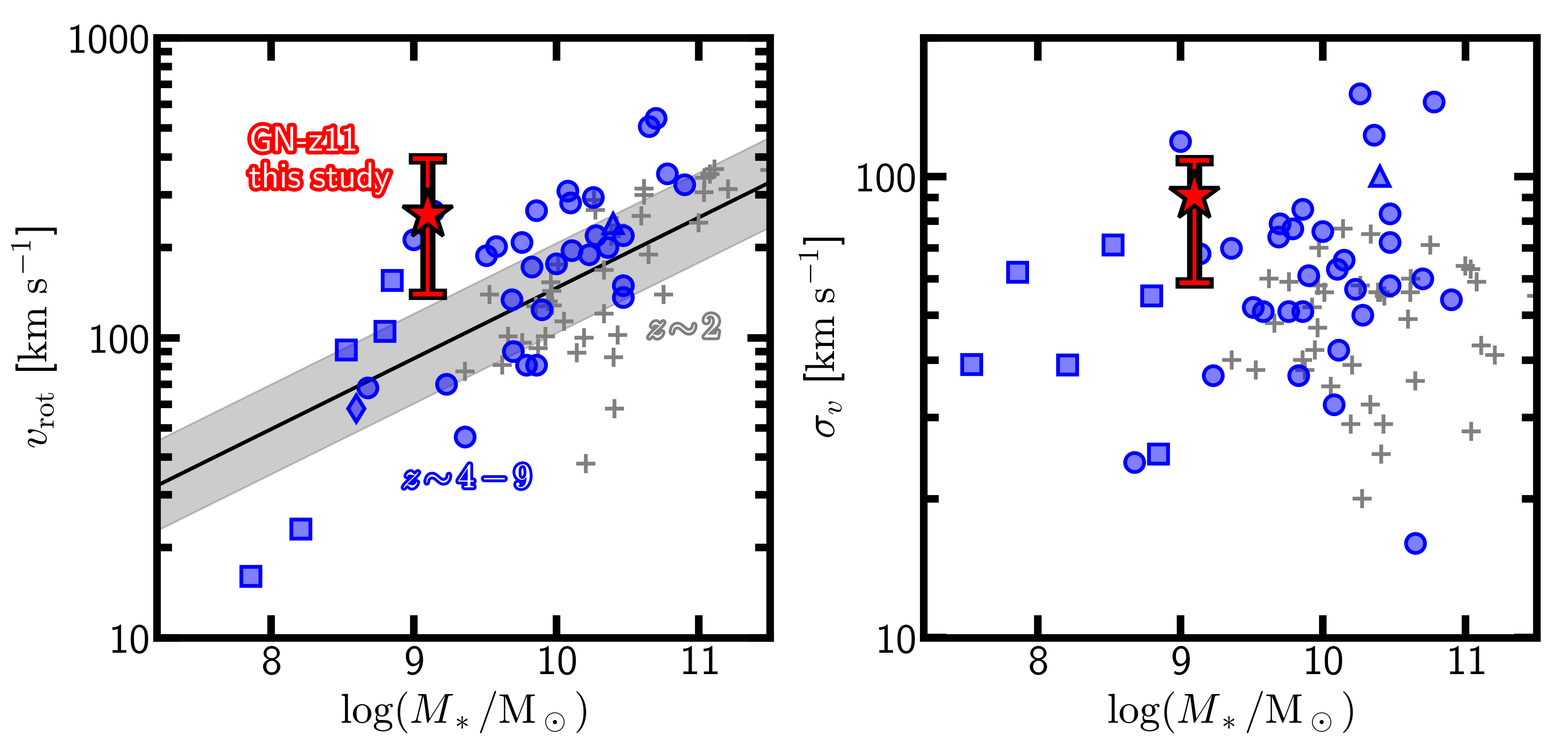}
    \caption{
        Rotation velocity and velocity dispersion as a function of stellar mass. The symbols are the same as those in Figure \ref{fig:voversigma_z}. The grey crosses are $z\sim2$ galaxies taken from \cite{ForsterSchreiber+18}. The grey line on the left panel indicates the Tully-Fisher relation calculated by \cite{DiTeodoro+21}.
    }
    \label{fig:mass_dependent}
\end{figure*}

In Figure \ref{fig:mass_dependent}, we examine mass dependence of the rotation velocity and velocity dispersion combining previous observations and our results of GN-z11. We adopt the stellar mass $\log(M_*/\msun)=9.1_{-0.4}^{+0.3}$ of GN-z11 derived by \cite{Tacchella+23}. GN-z11 shows comparable rotation velocity and velocity dispersion to some previous reported high-$z$ galaxies, but possibly larger than those with similar stellar mass of $10^9~\msun$. Since the \ciii\ doublets are only marginally resolved with moderate S/N and spectral resolution, we cannot rule out the possibility that the $\vrot$ and $\sigma_v$ of GN-z11 are overestimated even with our best efforts. Fortunately, GN-z11 will be observed with the $R\sim2700$ JWST NIRspec high resolution gratings in the coming cycle-3 missions. Nevertheless, it is possible that GN-z11 has $\vrot$ and $\sigma_v$ larger than those of $z\sim2$ star forming galaxies \citep{ForsterSchreiber+18} and offsetted from the local Tully-Fisher relation \citep[e.g.,][]{DiTeodoro+21}, which can be explained by the compactness of GN-z11. 

\begin{figure}[htb!]
    \centering
    \includegraphics[width=\linewidth]{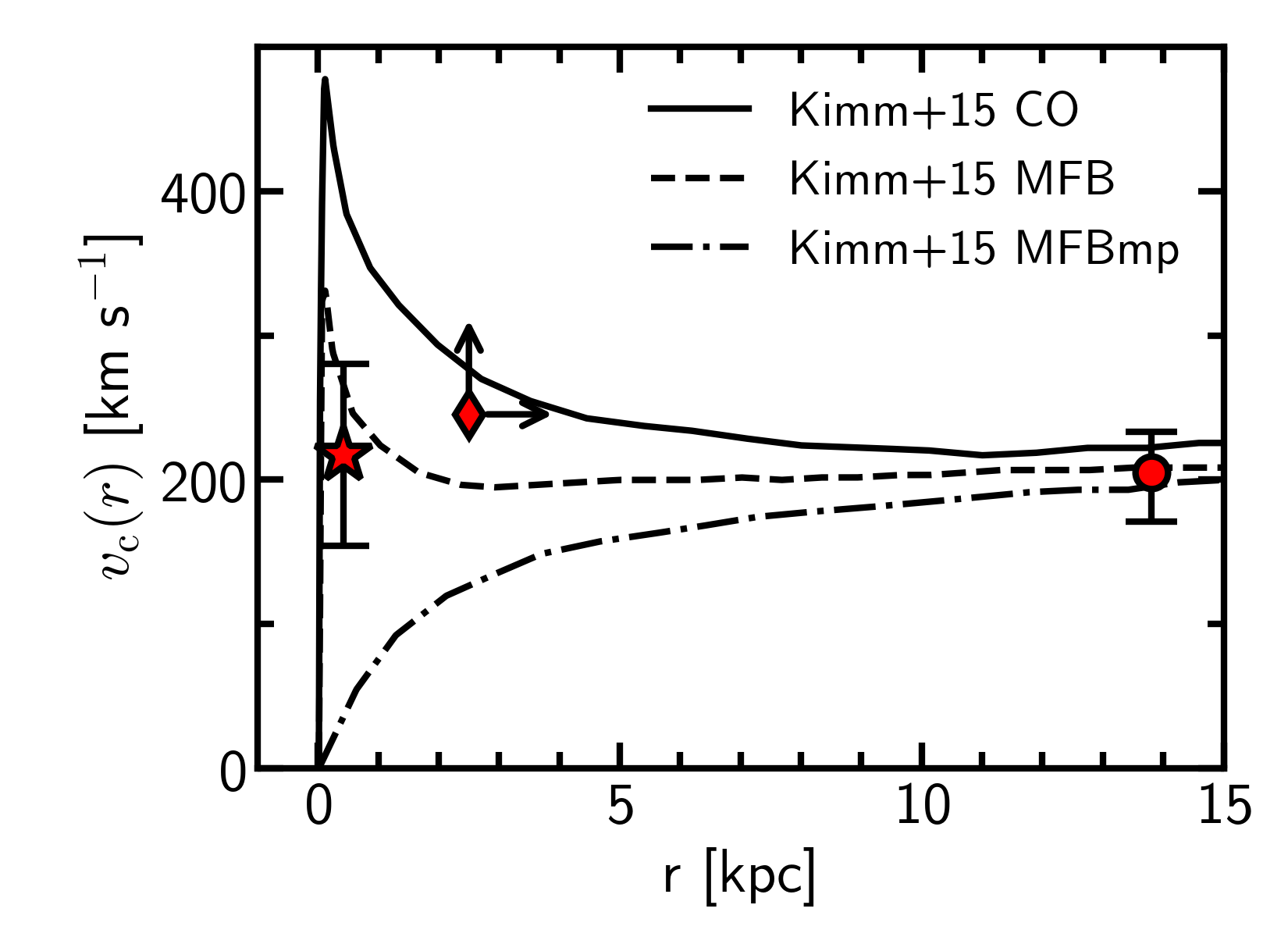}
    \caption{
    Circular velocities of GN-z11 at 
    $\rmsub{R}{e}$ (red star) and virial radius of the dark matter halo (red circle). \revise{Red diamond represents the offsetted He {\sc ii} clump identified by \cite{Maiolino+23b}, assuming its velocity shift originates from rotation around the center of GN-z11.} Black lines are the simulation predictions of \cite{Kimm+15} normalized \revise{to match the DM mass and halo radius of GN-z11. The notations CO, MFB, and MFBmp are taken from \cite{Kimm+15} representing no mechanical/kinetic feedback, mechanical feedback, and the most efficient case (mechanical feedback with porous ISM), respectively.}
    }
    \label{fig:vcir}
\end{figure}

\begin{figure}[htb!]
    \centering
    \includegraphics[width=\linewidth]{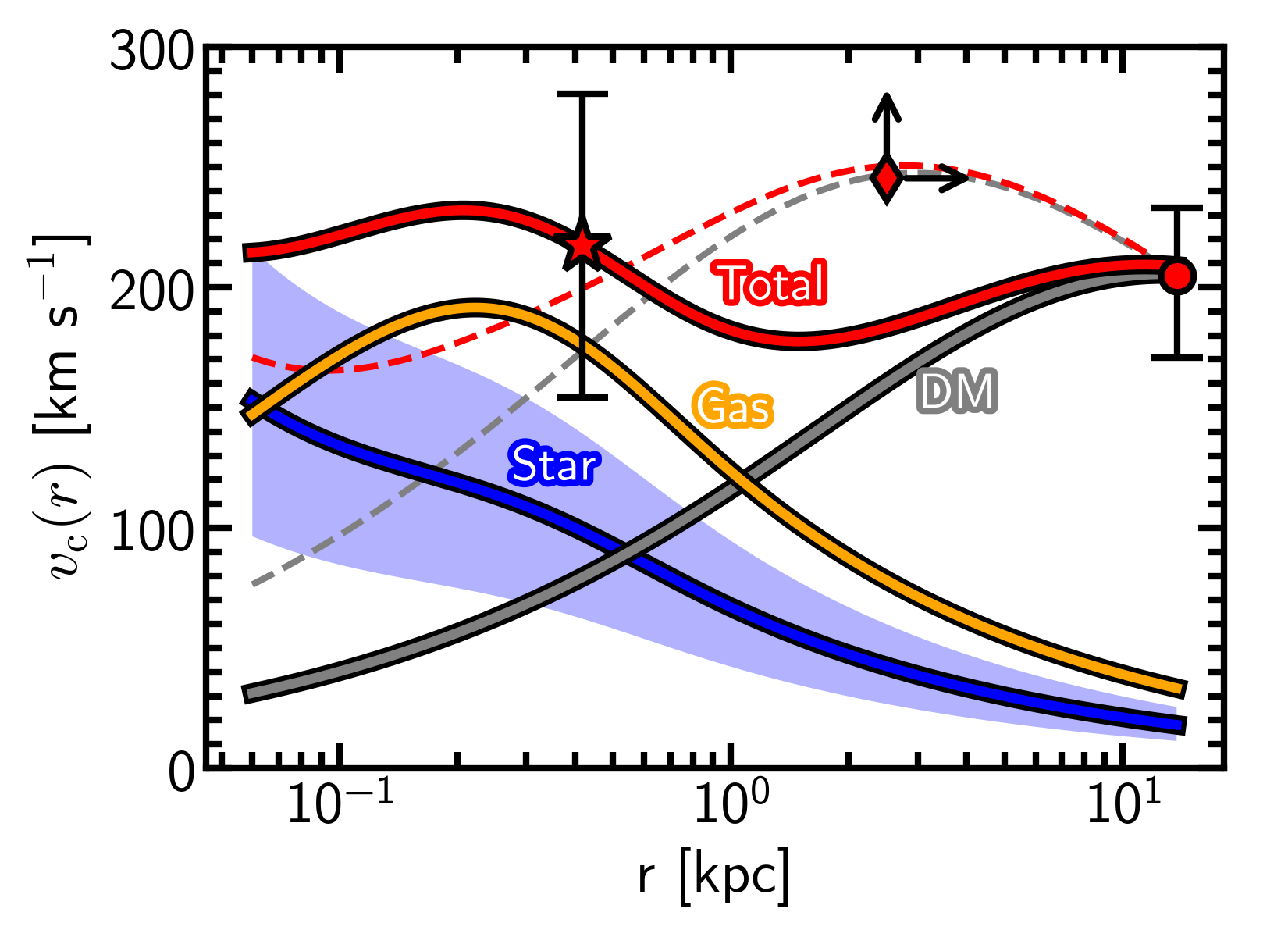}
    \caption{
    \revise{Data points are the same as those in Figure \ref{fig:vcir}. Solid lines are rotation curves given by different mass components, while the red line is the sum up. The DM mass is given by stellar-to-halo mass relation. For the stellar mass profile, we consider both the point-source and extended component identified by \cite{Tacchella+23}. Gas mass is the subtraction of stellar and DM mass from the dynamical mass at $2\rmsub{R}{e}$. Dashed grey line assumes the same DM mass but a concentrated distribution with $c=10$. The dashed red line including both the stellar mass and $c=10$ DM can also explain the observational data. }
    }
    \label{fig:vcir_budegt}
\end{figure}

To examine the mass concentration of GN-z11, we calculate the circular velocity from the rotation curve of Equation (\ref{eq:rotation_curve}) considering the asymmetric drift:
\begin{equation}
    \rmsub{v}{c}^2(r) = v(r)^2 + 2\sigma_v^2\times\left(\frac{r}{\rmsub{R}{d}}\right),
    \label{eq:vc_r}
\end{equation}
where \revise{$\rmsub{R}{d}=\rmsub{R}{e}/0.82$ is the radial scale length (see Equation (12) and (15) of \citealt{Burkert+16}). We obtain $\rmsub{v}{c}(2\rmsub{R}{e})=217\pm63~\kms$ at two times the effective radius} that roughly corresponds to the extension of \ciii\ investigated by this study.
Circular velocity probes the mass enclosed by $r$ with:
\begin{equation}
    v_\mathrm{c}(r)=\left(\frac{GM(<r)}{r}\right)^{1/2}.
    \label{eq:vc_r_mass}
\end{equation}
For the circular velocity of the DM halo, we can insert the halo mass ($\rmsub{M}{h}$) and the virial radius ($r_{200}$) using the following equation \citep{Mo&White02}:
\begin{equation}
    r_{200}=\left(\frac{GM_\mathrm{h}}{100\Omega_\mathrm{m}H_0^2}\right)^{1/3}(1+z)^{-1}.
\end{equation}
We adopt the $z\sim8$ stellar mass -- halo mass relation from \cite{Behroozi+19} and scale it to match the $z\sim10$ relation, as the stellar mass of GN-z11 is not well probed by the $z\sim10$ relation (see Figure 9 of \citealt{Behroozi+19}). We obtain $\rmsub{M}{h}=1.35_{-0.67}^{+0.56}\times10^{11}~\msun$ (c.f., \citealt{Scholtz+23,Ferrara23}) and $\rmsub{v}{c}(r_{200})=205_{-34}^{+28}~\kms$. \revise{As shown in Figure \ref{fig:vcir}, $\rmsub{v}{c}(2\rmsub{R}{e})$ is comparable to $\rmsub{v}{c}(r_{200})$ within the uncertainties. The compact size of GN-z11 thus would imply a steeply rising rotation curve with peaked velocity in the inner radius, which is similar to the model shown in \cite{Kimm+15} with no feedback (CO, solid line) or moderate mechanic feedback (MFB, dashed line) but distinguished from the one with effective feedback (MFBmp, dash-dotted line).} Such a centrally concentrated mass profile is a favorable explanation for disk formation and efficient star formation as predicted by simulations \citep[e.g.,][]{Hopkins+23}. \revise{Interestingly, the profile of $\rmsub{v}{c}(r)$ can be supported by the offsetted He {\sc ii} identified by \cite{Maiolino+23b}, although the velocity and spatial offset are measured as projected values and thus given as lower limits.}

\revise{In fact, the circular velocity profile can be further tested given the known size and mass measurements. In Figure \ref{fig:vcir_budegt}, we show the curve of circular velocity given by stellar mass. We adopt the two components identified by \cite{Tacchella+23}. For the point-source with $\log(M_*/\msun)=8.4_{-0.3}^{+0.3}$, we consider a simple $\rmsub{v}{c}\propto r^{-1}$ relation for $r>0''.06$. For the extended component with $\log(M_*/\msun)=8.9_{-0.3}^{+0.2}$,  we measure $\rmsub{v}{c}$ from the mass distribution of exponential profile. The shaded area shows the 1$\sigma$ uncertainty. For DM, we assume a Navarro–Frenk–White (NFW; \citealt{Navarro+96}) profile with a concentration parameter of $c=2.33$ following the $c-\rmsub{M}{h}$ relation of \cite{Dutton+14}. The contribution of DM is small in the inner region while the stellar mass alone cannot account for the disk rotation. There may be a significant mass budget from gas.}
\revise{From Equation (\ref{eq:vc_r_mass}), we obtain $M_\mathrm{dyn}(r<2\rmsub{R}{e})= 4.6\pm2.7\times10^9~\msun$. We then subtract the DM mass and stellar mass from dynamical mass to obtain a gas mass of $M_\mathrm{gas}(r<2\rmsub{R}{e})= 3.0\pm2.7\times10^9~\msun$. Including such a gas component and assuming the distribution of exponential disk with effective radius given by the fiducial disk model, we can explain the high rotation velocity as shown in Figure \ref{fig:vcir_budegt}. The gas fraction would be $\fgas\sim0.66$ within $2\rmsub{R}{e}$ suggestive of a gas-rich system.}
A similar gas mass fraction ($\fgas\sim0.5$) can be obtained if we adopt the SFR reported by \cite{Tacchella+23} and estimate the gas mass from the star formation rate surface density following the Kenicutt-Schmidt law \citep{Keinnicutt98}. 
\cite{Green+14} measure the gas mass for local star forming galaxies and find $M_*\sim10^9~\msun$ galaxies typically have $\fgas\sim0.5$.
Our estimation of gas fraction is interestingly comparable to those of local low-mass star forming galaxies, whose recent star formation is likely triggered by gas inflows \citep[e.g.,][]{Green+14,Isobe+23a,Xu+24}. 
\revise{We note that the contribution from DM could be more significant given the uncertainty on the DM profile. In Figure \ref{fig:vcir_budegt}, we also show the $\rmsub{v}{c}$ of DM assuming a concentrated profile with $c=10$ with dashed line. The combination of stellar mass and $c=10$ DM mass (red dashed line) is sufficient to explain the observed disk rotation. Higher spatial resolution is necessary to probe the detailed rotation curve and test different profiles of DM.}

\revise{Assuming GN-z11 is a gas rich system with the gas mass derived from dynamical mass, we obtain a high average gas density $n\sim1\times10^3~\mathrm{cm}^{-3}$. \cite{Dekel+23} suggest $n>5\times10^3~\mathrm{cm}^{-3}$ would allow star formation in disks under FFB.
We also find high surface density $\Sigma\sim10^4~\mathrm{M_\odot~pc^{-2}}$ suggestive of ineffective radiative feedback \citep{ZhaozhouLi+23}. As the density gets larger in the central region, GN-z11 would meet the criteria of FFB giving rise to its high UV luminosity. On the other hand, \cite{ZhaozhouLi+23} predict low $\fgas<0.1$ for galaxies residing in DM halos at the FFB threshold due to high star formation efficiency. The high $\fgas$ suggested by our results may indicate that feedback process is at play in a moderate manner or that GN-z11 is still undergoing starburst phase when the consumption of gas is incomplete.}

\revise{Rotational velocity, velocity dispersion, and gas fraction would provide a useful constraint on disk instability via the Toomre Q parameter: 
\begin{equation}
    Q=\frac{\sigma_v}{\vrot}\frac{a}{\fgas},
    \label{eq:ToomreQ}
\end{equation}
where the parameter $a$ ranges from 1 to 2 depending on the gas distribution. We assume $a=\sqrt{2}$ which corresponds to a disk with constant rotational velocity \citep[][]{Genzel+11}. We find $Q\approx1.32$ but with a large uncertainty $>1$, from which we cannot conclude $Q>0.67$ for stable thick disk. 
Note that accurate diagnostics of disk instability should also include multiple disk component (e.g., stellar component, ionized gas, neutral gas; see e.g., \citealt{Romeo+11,Romeo+13}) and the inhomogenous distribution of velocity dispersion \citep{Romeo+10,Romeo+14}. 
The current data is still insufficient to conclude the disk stability of GN-z11.}



\subsection{\revise{Outflows in GN-z11}}
\revise{Beside the rotating disk scenario, we check whether recent observational evidence can be consistently explained by galactic outflows. As discussed in Section \ref{analysis}, the extension of morphology in outskirt and the velocity gradient have an orientation roughly perpendicular to the extended component of GN-z11 identified by \cite{Tacchella+23}. The velocity difference of $\sim340~\kms$ (Table \ref{tab:results}) suggests an outflow velocity of $\sim170~\kms$ in the line-of-sight direction. Adopting an axial ratio ($b/a=0.67$) for the extended component, the inclination ($i$) can be estimated with $\cos i= (a^2/b^2-\alpha^2)/(1-\alpha^2)$, where $\alpha=0.3$ is the assumed disk height. We obtain $i=50^\circ$ and derive the deprojected outflow velocity of $\rmsub{v}{out}\sim270~\kms$. We emphasize that the outflow velocity is estimated based on the assumption that \ciii\ emission traces bi-conical outflows driven by the marginally resolved extended disk from NIRCam images. As a comparison, \cite{Bunker+23} find the Ly$\alpha$ emission has a velocity shift of $555~\kms$ from the systemic one suggestive of galactic outflows. \cite{Ferrara23} find the Ly$\alpha$ profile is consistent with a terminal outflow velocity of $200~\kms$ considering radiation-driven outflows, radiative transfer, and IGM attenuation. Because \ciii\ traces the ionized gas, our measurement agrees with the one derived by \cite{Ferrara23}. Deep NIRSpec G395H/F290LP or MIRI MRS observations of bright optical emission lines can further test the consistency of outflow velocity.}



\revise{We also recognise other alternative scenarios, such as galaxy mergers.}
Distinguishing between rotation disk and galaxy mergers has been a longstanding problem for high-$z$ galaxies in the case of limited spatial resolution and S/N \citep[see e.g.,][]{Rizzo+22}. GN-z11 is very compact and may not be well resolved until next-generation facilities such as extremely large telescopes with adaptive optics. With the current observational evidence, the scenario of major merger is not favored for GN-z11 and similar high$-z$ galaxies with compact sizes \citep[see e.g.,][]{Ono+23,Harikane+24b}.

\section{Summary}
\label{sum}
In this paper we investigate the dynamics of GN-z11, a luminous galaxy at $z=10.60$. We carefully analyze the public deep integral field spectroscopy (IFS) data taken with JWST NIRSpec IFU and exploit the spatial and spectral resolution to determine the dynamical properties of GN-z11. Our findings are summarized below:
\begin{itemize}
    \item We investigate the \ciii$\lambda\lambda1907,1909$ emission that traces ionized gas. The half-light radius of the \ciii\ emitting gas disk is only $221 \pm 42$ pc but spatially extended significantly beyond the PSF. \revise{The morphology of \ciii\ may be explained by an extended disk or galactic outflows.}
    \item The spatially extended \ciii\ emission of GN-z11 shows a velocity gradient, red- and blue-shifted components in the \revise{north and south} directions, respectively, which cannot be explained by the variation of [\ciii$\lambda$1907/\ciii$\lambda$1909 line ratios. We perform forward modeling with GalPak$^\mathrm{3D}$, including the effects of PSF smearing and line blending, and find that the best-fit \revise{rotating disk model has a rotation velocity of $v_\mathrm{rot}=257^{+138}_{-117}$ km s$^{-1}$ and a velocity dispersion of $\sigma_v=91^{+18}_{-32}$ km s$^{-1}$}. Both the observed kinematic ratio $\Delta v_\mathrm{obs}/2\sigma_\mathrm{med}=1.50$ and the intrinsic $v_\mathrm{rot}/\sigma_v=2.83^{+1.82}_{-1.41}$ would indicate a rotation-dominated disk. The rotation velocity and velocity dispersion of GN-z11 is comparable to $4<z<9$ galaxies previously reported by ground-based sub-millimeter observations and JWST, but possibly larger than those with similar stellar mass of $10^9~\msun$. Observations with better S/N and resolution are required to confirm this result.
    \item \revise{The disk rotation velocity becomes as fast as the circular velocity of DM halo ($\rmsub{v}{c}(r_{200})=205_{-34}^{+28}~\kms$) within a small radius,} suggesting a compact disk produced under weak feedback such predicted in numerical simulations. Our results would suggest a fast-rotating gaseous disk at $z>10$ whose center possesses luminous stellar components or AGN providing weak feedback. \revise{From the dynamical mass given by disk rotation, we estimate a high gas fraction and a high gas density, which can be compared to galaxy evolution models such as FFB.}
    \item \revise{An alternative explanation for the velocity gradient would be galactic outflows that extends in the polar direction of a disk component. We estimate an outflow velocity of $\sim280~\kms$ consistent other observational results for GN-z11 but requires further confirmation.}
\end{itemize}

This work is based on observations made with the NASA/ESA/CSA James Webb Space Telescope.
The data were obtained from the Mikulski Archive for Space Telescopes at the Space Telescope Science Institute, which is operated by the Association of Universities for Research in Astronomy, Inc., under NASA contract NAS 5-03127 for JWST.
These observations are associated with programs 1429, 1537, and 4426.
We thank the research team led by Roberto Maiolino for developing their observing program of GN-z11.
We thank Yuzo Ishikawa for the discussions on data analysis of JWST NIRspec IFU.
We thank Alessandro Romeo, Andrea Ferrara, Daniel Schaerer, Federico Lelli, Pascal Oesch, Seiji Fujimoto, Takafumi Tsukui, Toru Yamada for the useful discussions.
We are grateful to staff of the James Webb Space Telescope Help Desk for letting us know useful information.
This work was supported by the joint research program of the Institute for Cosmic Ray Research (ICRR), University of Tokyo. 
This publication is based upon work supported by the World Premier International Research Center Initiative (WPI Initiative), MEXT, Japan.
M.O., H.Y., H.F., K.N., Y.H., and Y.I. are supported by JSPS KAKENHI Grant Nos. 20H00180/21H04467, 21H04489, 3K13139, 20K22373, 21K13953, and 24KJ0202 respectively.
H.Y. is supported by JST FOREST Program, Grant Number JP-MJFR202Z.
\software{JWST Calibration Pipeline \citep{bushouse_2023_10022973}, astropy \citep{astropy:2013,astropy:2018,astropy:2022}, galfit \citep{Peng+02,Peng+10}, GalPak$^\mathrm{3D}$ \citep{Bouche+15}, WebbPFS \citep{WebbPSF}}

\bibliography{ref}{}

\begin{thebibliography}{}
\expandafter\ifx\csname natexlab\endcsname\relax\def\natexlab#1{#1}\fi
\providecommand{\url}[1]{\href{#1}{#1}}
\providecommand{\dodoi}[1]{doi:~\href{http://doi.org/#1}{\nolinkurl{#1}}}
\providecommand{\doeprint}[1]{\href{http://ascl.net/#1}{\nolinkurl{http://ascl.net/#1}}}
\providecommand{\doarXiv}[1]{\href{https://arxiv.org/abs/#1}{\nolinkurl{https://arxiv.org/abs/#1}}}

\bibitem[{{Arrabal Haro} {et~al.}(2023{\natexlab{a}}){Arrabal Haro},
  {Dickinson}, {Finkelstein}, {Fujimoto}, {Fern{\'a}ndez}, {Kartaltepe},
  {Jung}, {Cole}, {Burgarella}, {Chworowsky}, {Hutchison}, {Morales},
  {Papovich}, {Simons}, {Amor{\'\i}n}, {Backhaus}, {Bagley}, {Bisigello},
  {Calabr{\`o}}, {Castellano}, {Cleri}, {Dav{\'e}}, {Dekel}, {Ferguson},
  {Fontana}, {Gawiser}, {Giavalisco}, {Harish}, {Hathi}, {Hirschmann},
  {Holwerda}, {Huertas-Company}, {Koekemoer}, {Larson}, {Lucas}, {Mobasher},
  {P{\'e}rez-Gonz{\'a}lez}, {Pirzkal}, {Rose}, {Santini}, {Trump}, {de la
  Vega}, {Wang}, {Weiner}, {Wilkins}, {Yang}, {Yung}, \&
  {Zavala}}]{ArrabalHaro+23a}
{Arrabal Haro}, P., {Dickinson}, M., {Finkelstein}, S.~L., {et~al.}
  2023{\natexlab{a}}, \apjl, 951, L22, \dodoi{10.3847/2041-8213/acdd54}

\bibitem[{{Arrabal Haro} {et~al.}(2023{\natexlab{b}}){Arrabal Haro},
  {Dickinson}, {Finkelstein}, {Kartaltepe}, {Donnan}, {Burgarella}, {Carnall},
  {Cullen}, {Dunlop}, {Fern{\'a}ndez}, {Fujimoto}, {Jung}, {Krips}, {Larson},
  {Papovich}, {P{\'e}rez-Gonz{\'a}lez}, {Amor{\'\i}n}, {Bagley}, {Buat},
  {Casey}, {Chworowsky}, {Cohen}, {Ferguson}, {Giavalisco}, {Huertas-Company},
  {Hutchison}, {Kocevski}, {Koekemoer}, {Lucas}, {McLeod}, {McLure}, {Pirzkal},
  {Seill{\'e}}, {Trump}, {Weiner}, {Wilkins}, \& {Zavala}}]{ArrabalHaro+23b}
---. 2023{\natexlab{b}}, \nat, 622, 707, \dodoi{10.1038/s41586-023-06521-7}

\bibitem[{{Astropy Collaboration} {et~al.}(2013){Astropy Collaboration},
  {Robitaille}, {Tollerud}, {Greenfield}, {Droettboom}, {Bray}, {Aldcroft},
  {Davis}, {Ginsburg}, {Price-Whelan}, {Kerzendorf}, {Conley}, {Crighton},
  {Barbary}, {Muna}, {Ferguson}, {Grollier}, {Parikh}, {Nair}, {Unther},
  {Deil}, {Woillez}, {Conseil}, {Kramer}, {Turner}, {Singer}, {Fox}, {Weaver},
  {Zabalza}, {Edwards}, {Azalee Bostroem}, {Burke}, {Casey}, {Crawford},
  {Dencheva}, {Ely}, {Jenness}, {Labrie}, {Lim}, {Pierfederici}, {Pontzen},
  {Ptak}, {Refsdal}, {Servillat}, \& {Streicher}}]{astropy:2013}
{Astropy Collaboration}, {Robitaille}, T.~P., {Tollerud}, E.~J., {et~al.} 2013,
  \aap, 558, A33, \dodoi{10.1051/0004-6361/201322068}

\bibitem[{{Astropy Collaboration} {et~al.}(2018){Astropy Collaboration},
  {Price-Whelan}, {Sip{\H{o}}cz}, {G{\"u}nther}, {Lim}, {Crawford}, {Conseil},
  {Shupe}, {Craig}, {Dencheva}, {Ginsburg}, {Vand erPlas}, {Bradley},
  {P{\'e}rez-Su{\'a}rez}, {de Val-Borro}, {Aldcroft}, {Cruz}, {Robitaille},
  {Tollerud}, {Ardelean}, {Babej}, {Bach}, {Bachetti}, {Bakanov}, {Bamford},
  {Barentsen}, {Barmby}, {Baumbach}, {Berry}, {Biscani}, {Boquien}, {Bostroem},
  {Bouma}, {Brammer}, {Bray}, {Breytenbach}, {Buddelmeijer}, {Burke},
  {Calderone}, {Cano Rodr{\'\i}guez}, {Cara}, {Cardoso}, {Cheedella}, {Copin},
  {Corrales}, {Crichton}, {D'Avella}, {Deil}, {Depagne}, {Dietrich}, {Donath},
  {Droettboom}, {Earl}, {Erben}, {Fabbro}, {Ferreira}, {Finethy}, {Fox},
  {Garrison}, {Gibbons}, {Goldstein}, {Gommers}, {Greco}, {Greenfield},
  {Groener}, {Grollier}, {Hagen}, {Hirst}, {Homeier}, {Horton}, {Hosseinzadeh},
  {Hu}, {Hunkeler}, {Ivezi{\'c}}, {Jain}, {Jenness}, {Kanarek}, {Kendrew},
  {Kern}, {Kerzendorf}, {Khvalko}, {King}, {Kirkby}, {Kulkarni}, {Kumar},
  {Lee}, {Lenz}, {Littlefair}, {Ma}, {Macleod}, {Mastropietro}, {McCully},
  {Montagnac}, {Morris}, {Mueller}, {Mumford}, {Muna}, {Murphy}, {Nelson},
  {Nguyen}, {Ninan}, {N{\"o}the}, {Ogaz}, {Oh}, {Parejko}, {Parley}, {Pascual},
  {Patil}, {Patil}, {Plunkett}, {Prochaska}, {Rastogi}, {Reddy Janga},
  {Sabater}, {Sakurikar}, {Seifert}, {Sherbert}, {Sherwood-Taylor}, {Shih},
  {Sick}, {Silbiger}, {Singanamalla}, {Singer}, {Sladen}, {Sooley},
  {Sornarajah}, {Streicher}, {Teuben}, {Thomas}, {Tremblay}, {Turner},
  {Terr{\'o}n}, {van Kerkwijk}, {de la Vega}, {Watkins}, {Weaver}, {Whitmore},
  {Woillez}, {Zabalza}, \& {Astropy Contributors}}]{astropy:2018}
{Astropy Collaboration}, {Price-Whelan}, A.~M., {Sip{\H{o}}cz}, B.~M., {et~al.}
  2018, \aj, 156, 123, \dodoi{10.3847/1538-3881/aabc4f}

\bibitem[{{Astropy Collaboration} {et~al.}(2022){Astropy Collaboration},
  {Price-Whelan}, {Lim}, {Earl}, {Starkman}, {Bradley}, {Shupe}, {Patil},
  {Corrales}, {Brasseur}, {N{\"o}the}, {Donath}, {Tollerud}, {Morris},
  {Ginsburg}, {Vaher}, {Weaver}, {Tocknell}, {Jamieson}, {van Kerkwijk},
  {Robitaille}, {Merry}, {Bachetti}, {G{\"u}nther}, {Aldcroft},
  {Alvarado-Montes}, {Archibald}, {B{\'o}di}, {Bapat}, {Barentsen},
  {Baz{\'a}n}, {Biswas}, {Boquien}, {Burke}, {Cara}, {Cara}, {Conroy},
  {Conseil}, {Craig}, {Cross}, {Cruz}, {D'Eugenio}, {Dencheva}, {Devillepoix},
  {Dietrich}, {Eigenbrot}, {Erben}, {Ferreira}, {Foreman-Mackey}, {Fox},
  {Freij}, {Garg}, {Geda}, {Glattly}, {Gondhalekar}, {Gordon}, {Grant},
  {Greenfield}, {Groener}, {Guest}, {Gurovich}, {Handberg}, {Hart},
  {Hatfield-Dodds}, {Homeier}, {Hosseinzadeh}, {Jenness}, {Jones}, {Joseph},
  {Kalmbach}, {Karamehmetoglu}, {Ka{\l}uszy{\'n}ski}, {Kelley}, {Kern},
  {Kerzendorf}, {Koch}, {Kulumani}, {Lee}, {Ly}, {Ma}, {MacBride}, {Maljaars},
  {Muna}, {Murphy}, {Norman}, {O'Steen}, {Oman}, {Pacifici}, {Pascual},
  {Pascual-Granado}, {Patil}, {Perren}, {Pickering}, {Rastogi}, {Roulston},
  {Ryan}, {Rykoff}, {Sabater}, {Sakurikar}, {Salgado}, {Sanghi}, {Saunders},
  {Savchenko}, {Schwardt}, {Seifert-Eckert}, {Shih}, {Jain}, {Shukla}, {Sick},
  {Simpson}, {Singanamalla}, {Singer}, {Singhal}, {Sinha}, {Sip{\H{o}}cz},
  {Spitler}, {Stansby}, {Streicher}, {{\v{S}}umak}, {Swinbank}, {Taranu},
  {Tewary}, {Tremblay}, {de Val-Borro}, {Van Kooten}, {Vasovi{\'c}}, {Verma},
  {de Miranda Cardoso}, {Williams}, {Wilson}, {Winkel}, {Wood-Vasey}, {Xue},
  {Yoachim}, {Zhang}, {Zonca}, \& {Astropy Project
  Contributors}}]{astropy:2022}
{Astropy Collaboration}, {Price-Whelan}, A.~M., {Lim}, P.~L., {et~al.} 2022,
  \apj, 935, 167, \dodoi{10.3847/1538-4357/ac7c74}

\bibitem[{{Behroozi} {et~al.}(2019){Behroozi}, {Wechsler}, {Hearin}, \&
  {Conroy}}]{Behroozi+19}
{Behroozi}, P., {Wechsler}, R.~H., {Hearin}, A.~P., \& {Conroy}, C. 2019,
  \mnras, 488, 3143, \dodoi{10.1093/mnras/stz1182}

\bibitem[{{B{\"o}ker} {et~al.}(2022){B{\"o}ker}, {Arribas}, {L{\"u}tzgendorf},
  {Alves de Oliveira}, {Beck}, {Birkmann}, {Bunker}, {Charlot}, {de Marchi},
  {Ferruit}, {Giardino}, {Jakobsen}, {Kumari}, {L{\'o}pez-Caniego}, {Maiolino},
  {Manjavacas}, {Marston}, {Moseley}, {Muzerolle}, {Ogle}, {Pirzkal},
  {Rauscher}, {Rawle}, {Rix}, {Sabbi}, {Sargent}, {Sirianni}, {te Plate},
  {Valenti}, {Willott}, \& {Zeidler}}]{Beoker+22}
{B{\"o}ker}, T., {Arribas}, S., {L{\"u}tzgendorf}, N., {et~al.} 2022, \aap,
  661, A82, \dodoi{10.1051/0004-6361/202142589}

\bibitem[{{Bouch{\'e}} {et~al.}(2015){Bouch{\'e}}, {Carfantan}, {Schroetter},
  {Michel-Dansac}, \& {Contini}}]{Bouche+15}
{Bouch{\'e}}, N., {Carfantan}, H., {Schroetter}, I., {Michel-Dansac}, L., \&
  {Contini}, T. 2015, \aj, 150, 92, \dodoi{10.1088/0004-6256/150/3/92}

\bibitem[{{Bouwens} {et~al.}(2010){Bouwens}, {Illingworth}, {Gonz{\'a}lez},
  {Labb{\'e}}, {Franx}, {Conselice}, {Blakeslee}, {van Dokkum}, {Holden},
  {Magee}, {Marchesini}, \& {Zheng}}]{Bouwens+10}
{Bouwens}, R.~J., {Illingworth}, G.~D., {Gonz{\'a}lez}, V., {et~al.} 2010,
  \apj, 725, 1587, \dodoi{10.1088/0004-637X/725/2/1587}

\bibitem[{{Bunker} {et~al.}(2023){Bunker}, {Saxena}, {Cameron}, {Willott},
  {Curtis-Lake}, {Jakobsen}, {Carniani}, {Smit}, {Maiolino}, {Witstok},
  {Curti}, {D'Eugenio}, {Jones}, {Ferruit}, {Arribas}, {Charlot}, {Chevallard},
  {Giardino}, {de Graaff}, {Looser}, {L{\"u}tzgendorf}, {Maseda}, {Rawle},
  {Rix}, {Del Pino}, {Alberts}, {Egami}, {Eisenstein}, {Endsley}, {Hainline},
  {Hausen}, {Johnson}, {Rieke}, {Rieke}, {Robertson}, {Shivaei}, {Stark},
  {Sun}, {Tacchella}, {Tang}, {Williams}, {Willmer}, {Baker}, {Baum},
  {Bhatawdekar}, {Bowler}, {Boyett}, {Chen}, {Circosta}, {Helton}, {Ji},
  {Kumari}, {Lyu}, {Nelson}, {Parlanti}, {Perna}, {Sandles}, {Scholtz},
  {Suess}, {Topping}, {{\"U}bler}, {Wallace}, \& {Whitler}}]{Bunker+23}
{Bunker}, A.~J., {Saxena}, A., {Cameron}, A.~J., {et~al.} 2023, \aap, 677, A88,
  \dodoi{10.1051/0004-6361/202346159}

\bibitem[{{Burkert} {et~al.}(2016){Burkert}, {F{\"o}rster Schreiber}, {Genzel},
  {Lang}, {Tacconi}, {Wisnioski}, {Wuyts}, {Bandara}, {Beifiori}, {Bender},
  {Brammer}, {Chan}, {Davies}, {Dekel}, {Fabricius}, {Fossati}, {Kulkarni},
  {Lutz}, {Mendel}, {Momcheva}, {Nelson}, {Naab}, {Renzini}, {Saglia},
  {Sharples}, {Sternberg}, {Wilman}, \& {Wuyts}}]{Burkert+16}
{Burkert}, A., {F{\"o}rster Schreiber}, N.~M., {Genzel}, R., {et~al.} 2016,
  \apj, 826, 214, \dodoi{10.3847/0004-637X/826/2/214}

\bibitem[{Bushouse {et~al.}(2023)Bushouse, Eisenhamer, Dencheva, Davies,
  Greenfield, Morrison, Hodge, Simon, Grumm, Droettboom, Slavich, Sosey, Pauly,
  Miller, Jedrzejewski, Hack, Davis, Crawford, Law, Gordon, Regan, Cara,
  MacDonald, Bradley, Shanahan, Jamieson, Teodoro, Williams, \&
  Pena-Guerrero}]{bushouse_2023_10022973}
Bushouse, H., Eisenhamer, J., Dencheva, N., {et~al.} 2023, JWST Calibration
  Pipeline, 1.12.5,  Zenodo, \dodoi{10.5281/zenodo.10022973}

\bibitem[{{Curtis-Lake} {et~al.}(2023){Curtis-Lake}, {Carniani}, {Cameron},
  {Charlot}, {Jakobsen}, {Maiolino}, {Bunker}, {Witstok}, {Smit}, {Chevallard},
  {Willott}, {Ferruit}, {Arribas}, {Bonaventura}, {Curti}, {D'Eugenio},
  {Franx}, {Giardino}, {Looser}, {L{\"u}tzgendorf}, {Maseda}, {Rawle}, {Rix},
  {Rodr{\'\i}guez del Pino}, {{\"U}bler}, {Sirianni}, {Dressler}, {Egami},
  {Eisenstein}, {Endsley}, {Hainline}, {Hausen}, {Johnson}, {Rieke},
  {Robertson}, {Shivaei}, {Stark}, {Tacchella}, {Williams}, {Willmer},
  {Bhatawdekar}, {Bowler}, {Boyett}, {Chen}, {de Graaff}, {Helton}, {Hviding},
  {Jones}, {Kumari}, {Lyu}, {Nelson}, {Perna}, {Sandles}, {Saxena}, {Suess},
  {Sun}, {Topping}, {Wallace}, \& {Whitler}}]{CurtisLake+23}
{Curtis-Lake}, E., {Carniani}, S., {Cameron}, A., {et~al.} 2023, Nature
  Astronomy, 7, 622, \dodoi{10.1038/s41550-023-01918-w}

\bibitem[{{de Blok} {et~al.}(2008){de Blok}, {Walter}, {Brinks},
  {Trachternach}, {Oh}, \& {Kennicutt}}]{deBlok+08}
{de Blok}, W.~J.~G., {Walter}, F., {Brinks}, E., {et~al.} 2008, \aj, 136, 2648,
  \dodoi{10.1088/0004-6256/136/6/2648}

\bibitem[{{de Graaff} {et~al.}(2023){de Graaff}, {Rix}, {Carniani}, {Suess},
  {Charlot}, {Curtis-Lake}, {Arribas}, {Baker}, {Boyett}, {Bunker}, {Cameron},
  {Chevallard}, {Curti}, {Eisenstein}, {Franx}, {Hainline}, {Hausen}, {Ji},
  {Johnson}, {Jones}, {Maiolino}, {Maseda}, {Nelson}, {Parlanti}, {Rawle},
  {Robertson}, {Tacchella}, {{\"U}bler}, {Williams}, {Willmer}, \&
  {Willott}}]{deGraaf+23}
{de Graaff}, A., {Rix}, H.-W., {Carniani}, S., {et~al.} 2023, arXiv e-prints,
  arXiv:2308.09742, \dodoi{10.48550/arXiv.2308.09742}

\bibitem[{{Dekel} {et~al.}(2009){Dekel}, {Sari}, \& {Ceverino}}]{Dekel+09}
{Dekel}, A., {Sari}, R., \& {Ceverino}, D. 2009, \apj, 703, 785,
  \dodoi{10.1088/0004-637X/703/1/785}

\bibitem[{{Dekel} {et~al.}(2023){Dekel}, {Sarkar}, {Birnboim}, {Mandelker}, \&
  {Li}}]{Dekel+23}
{Dekel}, A., {Sarkar}, K.~C., {Birnboim}, Y., {Mandelker}, N., \& {Li}, Z.
  2023, \mnras, 523, 3201, \dodoi{10.1093/mnras/stad1557}

\bibitem[{{D'Eugenio} {et~al.}(2023){D'Eugenio}, {Perez-Gonzalez}, {Maiolino},
  {Scholtz}, {Perna}, {Circosta}, {Uebler}, {Arribas}, {Boeker}, {Bunker},
  {Carniani}, {Charlot}, {Chevallard}, {Cresci}, {Curtis-Lake}, {Jones},
  {Kumari}, {Lamperti}, {Looser}, {Parlanti}, {Rix}, {Robertson}, {Rodriguez
  Del Pino}, {Tacchella}, {Venturi}, \& {Willott}}]{DEugenio+23}
{D'Eugenio}, F., {Perez-Gonzalez}, P., {Maiolino}, R., {et~al.} 2023, arXiv
  e-prints, arXiv:2308.06317, \dodoi{10.48550/arXiv.2308.06317}

\bibitem[{{Di Teodoro} {et~al.}(2021){Di Teodoro}, {Posti}, {Ogle}, {Fall}, \&
  {Jarrett}}]{DiTeodoro+21}
{Di Teodoro}, E.~M., {Posti}, L., {Ogle}, P.~M., {Fall}, S.~M., \& {Jarrett},
  T. 2021, \mnras, 507, 5820, \dodoi{10.1093/mnras/stab2549}

\bibitem[{{Donnan} {et~al.}(2023){Donnan}, {McLeod}, {Dunlop}, {McLure},
  {Carnall}, {Begley}, {Cullen}, {Hamadouche}, {Bowler}, {Magee}, {McCracken},
  {Milvang-Jensen}, {Moneti}, \& {Targett}}]{Donnan+23a}
{Donnan}, C.~T., {McLeod}, D.~J., {Dunlop}, J.~S., {et~al.} 2023, \mnras, 518,
  6011, \dodoi{10.1093/mnras/stac3472}

\bibitem[{{Dutton} \& {Macci{\`o}}(2014)}]{Dutton+14}
{Dutton}, A.~A., \& {Macci{\`o}}, A.~V. 2014, \mnras, 441, 3359,
  \dodoi{10.1093/mnras/stu742}

\bibitem[{{Ferrara}(2023)}]{Ferrara23}
{Ferrara}, A. 2023, arXiv e-prints, arXiv:2310.12197,
  \dodoi{10.48550/arXiv.2310.12197}

\bibitem[{{F{\"o}rster Schreiber} {et~al.}(2009){F{\"o}rster Schreiber},
  {Genzel}, {Bouch{\'e}}, {Cresci}, {Davies}, {Buschkamp}, {Shapiro},
  {Tacconi}, {Hicks}, {Genel}, {Shapley}, {Erb}, {Steidel}, {Lutz},
  {Eisenhauer}, {Gillessen}, {Sternberg}, {Renzini}, {Cimatti}, {Daddi},
  {Kurk}, {Lilly}, {Kong}, {Lehnert}, {Nesvadba}, {Verma}, {McCracken},
  {Arimoto}, {Mignoli}, \& {Onodera}}]{ForsterSchreiber+09}
{F{\"o}rster Schreiber}, N.~M., {Genzel}, R., {Bouch{\'e}}, N., {et~al.} 2009,
  \apj, 706, 1364, \dodoi{10.1088/0004-637X/706/2/1364}

\bibitem[{{F{\"o}rster Schreiber} {et~al.}(2018){F{\"o}rster Schreiber},
  {Renzini}, {Mancini}, {Genzel}, {Bouch{\'e}}, {Cresci}, {Hicks}, {Lilly},
  {Peng}, {Burkert}, {Carollo}, {Cimatti}, {Daddi}, {Davies}, {Genel}, {Kurk},
  {Lang}, {Lutz}, {Mainieri}, {McCracken}, {Mignoli}, {Naab}, {Oesch},
  {Pozzetti}, {Scodeggio}, {Shapiro Griffin}, {Shapley}, {Sternberg},
  {Tacchella}, {Tacconi}, {Wuyts}, \& {Zamorani}}]{ForsterSchreiber+18}
{F{\"o}rster Schreiber}, N.~M., {Renzini}, A., {Mancini}, C., {et~al.} 2018,
  \apjs, 238, 21, \dodoi{10.3847/1538-4365/aadd49}

\bibitem[{{Fraternali} {et~al.}(2021){Fraternali}, {Karim}, {Magnelli},
  {G{\'o}mez-Guijarro}, {Jim{\'e}nez-Andrade}, \& {Posses}}]{Fraternali+21}
{Fraternali}, F., {Karim}, A., {Magnelli}, B., {et~al.} 2021, \aap, 647, A194,
  \dodoi{10.1051/0004-6361/202039807}

\bibitem[{{Freeman}(1970)}]{Freeman70}
{Freeman}, K.~C. 1970, \apj, 160, 811, \dodoi{10.1086/150474}

\bibitem[{{Fujimoto} {et~al.}(2024){Fujimoto}, {Ouchi}, {Kohno}, {Valentino},
  {Gim{\'e}nez-Arteaga}, {Brammer}, {Furtak}, {Kohandel}, {Oguri},
  {Pallottini}, {Richard}, {Zitrin}, {Bauer}, {Boylan-Kolchin},
  {Dessauges-Zavadsky}, {Egami}, {Finkelstein}, {Ma}, {Smail}, {Watson},
  {Hutchison}, {Rigby}, {Welch}, {Ao}, {Bradley}, {Caminha}, {Caputi},
  {Espada}, {Endsley}, {Fudamoto}, {Gonz{\'a}lez-L{\'o}pez}, {Hatsukade},
  {Koekemoer}, {Kokorev}, {Laporte}, {Lee}, {Magdis}, {Ono}, {Rizzo},
  {Shibuya}, {Shimasaku}, {Sun}, {Toft}, {Umehata}, {Wang}, \&
  {Yajima}}]{Fujimoto+24}
{Fujimoto}, S., {Ouchi}, M., {Kohno}, K., {et~al.} 2024, arXiv e-prints,
  arXiv:2402.18543, \dodoi{10.48550/arXiv.2402.18543}

\bibitem[{{Genzel} {et~al.}(2011){Genzel}, {Newman}, {Jones}, {F{\"o}rster
  Schreiber}, {Shapiro}, {Genel}, {Lilly}, {Renzini}, {Tacconi}, {Bouch{\'e}},
  {Burkert}, {Cresci}, {Buschkamp}, {Carollo}, {Ceverino}, {Davies}, {Dekel},
  {Eisenhauer}, {Hicks}, {Kurk}, {Lutz}, {Mancini}, {Naab}, {Peng},
  {Sternberg}, {Vergani}, \& {Zamorani}}]{Genzel+11}
{Genzel}, R., {Newman}, S., {Jones}, T., {et~al.} 2011, \apj, 733, 101,
  \dodoi{10.1088/0004-637X/733/2/101}

\bibitem[{{Green} {et~al.}(2014){Green}, {Glazebrook}, {McGregor}, {Damjanov},
  {Wisnioski}, {Abraham}, {Colless}, {Sharp}, {Crain}, {Poole}, \&
  {McCarthy}}]{Green+14}
{Green}, A.~W., {Glazebrook}, K., {McGregor}, P.~J., {et~al.} 2014, \mnras,
  437, 1070, \dodoi{10.1093/mnras/stt1882}

\bibitem[{{Harikane} {et~al.}(2024{\natexlab{a}}){Harikane}, {Nakajima},
  {Ouchi}, {Umeda}, {Isobe}, {Ono}, {Xu}, \& {Zhang}}]{Harikane+24}
{Harikane}, Y., {Nakajima}, K., {Ouchi}, M., {et~al.} 2024{\natexlab{a}}, \apj,
  960, 56, \dodoi{10.3847/1538-4357/ad0b7e}

\bibitem[{{Harikane} {et~al.}(2023){Harikane}, {Ouchi}, {Oguri}, {Ono},
  {Nakajima}, {Isobe}, {Umeda}, {Mawatari}, \& {Zhang}}]{Harikane+23a}
{Harikane}, Y., {Ouchi}, M., {Oguri}, M., {et~al.} 2023, \apjs, 265, 5,
  \dodoi{10.3847/1538-4365/acaaa9}

\bibitem[{{Harikane} {et~al.}(2024{\natexlab{b}}){Harikane}, {Inoue}, {Ellis},
  {Ouchi}, {Nakazato}, {Yoshida}, {Ono}, {Sun}, {Sato}, {Fujimoto},
  {Kashikawa}, {McLeod}, {Perez-Gonzalez}, {Sawicki}, {Sugahara}, {Xu},
  {Yamanaka}, {Carnall}, {Cullen}, {Dunlop}, {Egami}, {Grogin}, {Isobe},
  {Koekemoer}, {Laporte}, {Lee}, {Magee}, {Matsuo}, {Matsuoka}, {Mawatari},
  {Nakajima}, {Nakane}, {Tamura}, {Umeda}, \& {Yanagisawa}}]{Harikane+24b}
{Harikane}, Y., {Inoue}, A.~K., {Ellis}, R.~S., {et~al.} 2024{\natexlab{b}},
  arXiv e-prints, arXiv:2406.18352, \dodoi{10.48550/arXiv.2406.18352}

\bibitem[{{Herrera-Camus} {et~al.}(2022){Herrera-Camus}, {F{\"o}rster
  Schreiber}, {Price}, {{\"U}bler}, {Bolatto}, {Davies}, {Fisher}, {Genzel},
  {Lutz}, {Naab}, {Nestor}, {Shimizu}, {Sternberg}, {Tacconi}, \&
  {Tadaki}}]{Herrera-Camus+22}
{Herrera-Camus}, R., {F{\"o}rster Schreiber}, N.~M., {Price}, S.~H., {et~al.}
  2022, \aap, 665, L8, \dodoi{10.1051/0004-6361/202142562}

\bibitem[{{Hopkins} {et~al.}(2012){Hopkins}, {Kere{\v{s}}}, {Murray},
  {Quataert}, \& {Hernquist}}]{Hopkins+12}
{Hopkins}, P.~F., {Kere{\v{s}}}, D., {Murray}, N., {Quataert}, E., \&
  {Hernquist}, L. 2012, \mnras, 427, 968,
  \dodoi{10.1111/j.1365-2966.2012.21981.x}

\bibitem[{{Hopkins} {et~al.}(2023){Hopkins}, {Gurvich}, {Shen}, {Hafen},
  {Grudi{\'c}}, {Kurinchi-Vendhan}, {Hayward}, {Jiang}, {Orr}, {Wetzel},
  {Kere{\v{s}}}, {Stern}, {Faucher-Gigu{\`e}re}, {Bullock}, {Wheeler},
  {El-Badry}, {Loebman}, {Moreno}, {Boylan-Kolchin}, \&
  {Quataert}}]{Hopkins+23}
{Hopkins}, P.~F., {Gurvich}, A.~B., {Shen}, X., {et~al.} 2023, \mnras, 525,
  2241, \dodoi{10.1093/mnras/stad1902}

\bibitem[{{Isobe} {et~al.}(2023{\natexlab{a}}){Isobe}, {Ouchi}, {Nakajima},
  {Harikane}, {Ono}, {Xu}, {Zhang}, \& {Umeda}}]{Isobe+23b}
{Isobe}, Y., {Ouchi}, M., {Nakajima}, K., {et~al.} 2023{\natexlab{a}}, \apj,
  956, 139, \dodoi{10.3847/1538-4357/acf376}

\bibitem[{{Isobe} {et~al.}(2023{\natexlab{b}}){Isobe}, {Ouchi}, {Nakajima},
  {Ozaki}, {Bouch{\'e}}, {Wise}, {Xu}, {Emsellem}, {Kusakabe}, {Hattori},
  {Nagao}, {Chiaki}, {Fukushima}, {Harikane}, {Hayashi}, {Hirai}, {Kim},
  {Maseda}, {Nagamine}, {Shibuya}, {Sugahara}, {Yajima}, {Aoyama}, {Fujimoto},
  {Fukushima}, {Hatano}, {Inoue}, {Ishigaki}, {Kawasaki}, {Kojima}, {Komiyama},
  {Koyama}, {Koyama}, {Lee}, {Matsumoto}, {Mawatari}, {Moriya}, {Motohara},
  {Murai}, {Nishigaki}, {Onodera}, {Ono}, {Rauch}, {Saito}, {Sasaki}, {Suzuki},
  {Takeuchi}, {Umeda}, {Umemura}, {Watanabe}, {Yabe}, \& {Zhang}}]{Isobe+23a}
---. 2023{\natexlab{b}}, \apj, 951, 102, \dodoi{10.3847/1538-4357/accc87}

\bibitem[{{Jakobsen} {et~al.}(2022){Jakobsen}, {Ferruit}, {Alves de Oliveira},
  {Arribas}, {Bagnasco}, {Barho}, {Beck}, {Birkmann}, {B{\"o}ker}, {Bunker},
  {Charlot}, {de Jong}, {de Marchi}, {Ehrenwinkler}, {Falcolini}, {Fels},
  {Franx}, {Franz}, {Funke}, {Giardino}, {Gnata}, {Holota}, {Honnen}, {Jensen},
  {Jentsch}, {Johnson}, {Jollet}, {Karl}, {Kling}, {K{\"o}hler}, {Kolm},
  {Kumari}, {Lander}, {Lemke}, {L{\'o}pez-Caniego}, {L{\"u}tzgendorf},
  {Maiolino}, {Manjavacas}, {Marston}, {Maschmann}, {Maurer}, {Messerschmidt},
  {Moseley}, {Mosner}, {Mott}, {Muzerolle}, {Pirzkal}, {Pittet}, {Plitzke},
  {Posselt}, {Rapp}, {Rauscher}, {Rawle}, {Rix}, {R{\"o}del}, {Rumler},
  {Sabbi}, {Salvignol}, {Schmid}, {Sirianni}, {Smith}, {Strada}, {te Plate},
  {Valenti}, {Wettemann}, {Wiehe}, {Wiesmayer}, {Willott}, {Wright}, {Zeidler},
  \& {Zincke}}]{Jakobeson+22}
{Jakobsen}, P., {Ferruit}, P., {Alves de Oliveira}, C., {et~al.} 2022, \aap,
  661, A80, \dodoi{10.1051/0004-6361/202142663}

\bibitem[{{Kennicutt}(1998)}]{Keinnicutt98}
{Kennicutt}, Robert~C., J. 1998, \apj, 498, 541, \dodoi{10.1086/305588}

\bibitem[{{Kewley} {et~al.}(2019){Kewley}, {Nicholls}, {Sutherland}, {Rigby},
  {Acharya}, {Dopita}, \& {Bayliss}}]{Kewley+19}
{Kewley}, L.~J., {Nicholls}, D.~C., {Sutherland}, R., {et~al.} 2019, \apj, 880,
  16, \dodoi{10.3847/1538-4357/ab16ed}

\bibitem[{{Kimm} {et~al.}(2015){Kimm}, {Cen}, {Devriendt}, {Dubois}, \&
  {Slyz}}]{Kimm+15}
{Kimm}, T., {Cen}, R., {Devriendt}, J., {Dubois}, Y., \& {Slyz}, A. 2015,
  \mnras, 451, 2900, \dodoi{10.1093/mnras/stv1211}

\bibitem[{{Li} {et~al.}(2023{\natexlab{a}}){Li}, {Dekel}, {Sarkar}, {Aung},
  {Giavalisco}, {Mandelker}, \& {Tacchella}}]{ZhaozhouLi+23}
{Li}, Z., {Dekel}, A., {Sarkar}, K.~C., {et~al.} 2023{\natexlab{a}}, arXiv
  e-prints, arXiv:2311.14662, \dodoi{10.48550/arXiv.2311.14662}

\bibitem[{{Li} {et~al.}(2023{\natexlab{b}}){Li}, {Cai}, {Sun}, {Richard},
  {Trebitsch}, {Helton}, {Diego}, {Oguri}, {Foo}, {Lin}, {Bauer}, {Chen},
  {Conselice}, {Espada}, {Egami}, {Fan}, {Frye}, {Fudamoto}, {Perez-Gonzalez},
  {Hainline}, {Hsiao}, {Ji}, {Jin}, {Koekemoer}, {Kokorev}, {Kohno}, {Li},
  {Lee}, {Magdis}, {Willmer}, {Windhorst}, {Wu}, {Yan}, {Zhang}, {Zitrin},
  {Zou}, {Bian}, {Cheng}, {DeCoursey}, {Furtak}, {Steinhardt}, \&
  {Umehata}}]{Li+23}
{Li}, Z., {Cai}, Z., {Sun}, F., {et~al.} 2023{\natexlab{b}}, arXiv e-prints,
  arXiv:2310.09327, \dodoi{10.48550/arXiv.2310.09327}

\bibitem[{{Maiolino} {et~al.}(2023){Maiolino}, {Uebler}, {Perna}, {Scholtz},
  {D'Eugenio}, {Witten}, {Laporte}, {Witstok}, {Carniani}, {Tacchella},
  {Baker}, {Arribas}, {Nakajima}, {Eisenstein}, {Bunker}, {Charlot}, {Cresci},
  {Curti}, {Curtis-Lake}, {de Graaff}, {Ji}, {Johnson}, {Kumari}, {Looser},
  {Maseda}, {Robertson}, {Rodriguez Del Pino}, {Sandles}, {Simmonds}, {Smit},
  {Sun}, {Venturi}, {Williams}, \& {Willmer}}]{Maiolino+23b}
{Maiolino}, R., {Uebler}, H., {Perna}, M., {et~al.} 2023, arXiv e-prints,
  arXiv:2306.00953, \dodoi{10.48550/arXiv.2306.00953}

\bibitem[{{Maiolino} {et~al.}(2024{\natexlab{a}}){Maiolino}, {Scholtz},
  {Witstok}, {Carniani}, {D'Eugenio}, {de Graaff}, {{\"U}bler}, {Tacchella},
  {Curtis-Lake}, {Arribas}, {Bunker}, {Charlot}, {Chevallard}, {Curti},
  {Looser}, {Maseda}, {Rawle}, {Rodr{\'\i}guez del Pino}, {Willott}, {Egami},
  {Eisenstein}, {Hainline}, {Robertson}, {Williams}, {Willmer}, {Baker},
  {Boyett}, {DeCoursey}, {Fabian}, {Helton}, {Ji}, {Jones}, {Kumari},
  {Laporte}, {Nelson}, {Perna}, {Sandles}, {Shivaei}, \& {Sun}}]{Maiolino+24}
{Maiolino}, R., {Scholtz}, J., {Witstok}, J., {et~al.} 2024{\natexlab{a}},
  \nat, 627, 59, \dodoi{10.1038/s41586-024-07052-5}

\bibitem[{{Maiolino} {et~al.}(2024{\natexlab{b}}){Maiolino}, {Risaliti},
  {Signorini}, {Trefoloni}, {Juodzbalis}, {Scholtz}, {Uebler}, {D'Eugenio},
  {Carniani}, {Fabian}, {Ji}, {Mazzolari}, {Bertola}, {Brusa}, {Bunker},
  {Charlot}, {Comastri}, {Cresci}, {DeCoursey}, {Egami}, {Fiore}, {Gilli},
  {Perna}, {Tacchella}, \& {Venturi}}]{Maiolino+24b}
{Maiolino}, R., {Risaliti}, G., {Signorini}, M., {et~al.} 2024{\natexlab{b}},
  arXiv e-prints, arXiv:2405.00504, \dodoi{10.48550/arXiv.2405.00504}

\bibitem[{{Mo} \& {White}(2002)}]{Mo&White02}
{Mo}, H.~J., \& {White}, S.~D.~M. 2002, \mnras, 336, 112,
  \dodoi{10.1046/j.1365-8711.2002.05723.x}

\bibitem[{{Naab} \& {Ostriker}(2017)}]{Naab+17}
{Naab}, T., \& {Ostriker}, J.~P. 2017, \araa, 55, 59,
  \dodoi{10.1146/annurev-astro-081913-040019}

\bibitem[{{Navarro} {et~al.}(1996){Navarro}, {Frenk}, \& {White}}]{Navarro+96}
{Navarro}, J.~F., {Frenk}, C.~S., \& {White}, S. D.~M. 1996, \apj, 462, 563,
  \dodoi{10.1086/177173}

\bibitem[{{Neeleman} {et~al.}(2020){Neeleman}, {Prochaska}, {Kanekar}, \&
  {Rafelski}}]{Neeleman+20}
{Neeleman}, M., {Prochaska}, J.~X., {Kanekar}, N., \& {Rafelski}, M. 2020,
  \nat, 581, 269, \dodoi{10.1038/s41586-020-2276-y}

\bibitem[{{Nelson} {et~al.}(2019){Nelson}, {Pillepich}, {Springel}, {Pakmor},
  {Weinberger}, {Genel}, {Torrey}, {Vogelsberger}, {Marinacci}, \&
  {Hernquist}}]{Nelson+19}
{Nelson}, D., {Pillepich}, A., {Springel}, V., {et~al.} 2019, \mnras, 490,
  3234, \dodoi{10.1093/mnras/stz2306}

\bibitem[{{Nelson} {et~al.}(2023){Nelson}, {Brammer}, {Gimenez-Arteaga},
  {Oesch}, {Ubler}, {de Graaff}, {Matharu}, {Naidu}, {Shapley}, {Whitaker},
  {Wisnioski}, {Forster Schreiber}, {Smit}, {van Dokkum}, {Chisholm},
  {Endsley}, {Hartley}, {Gibson}, {Giovinazzo}, {Illingworth}, {Labbe},
  {Maseda}, {Matthee}, {Covelo Paz}, {Price}, {Reddy}, {Shivaei}, {Weibel},
  {Wuyts}, {Xiao}, {Alberts}, {Baker}, {Bunker}, {Cameron}, {Charlot},
  {Eisenstein}, {Ji}, {Johnson}, {Jones}, {Maiolino}, {Robertson}, {Sandles},
  {Suess}, {Tacchella}, {Williams}, \& {Witstok}}]{Nelson+23}
{Nelson}, E.~J., {Brammer}, G., {Gimenez-Arteaga}, C., {et~al.} 2023, arXiv
  e-prints, arXiv:2310.06887, \dodoi{10.48550/arXiv.2310.06887}

\bibitem[{{Oesch} {et~al.}(2016){Oesch}, {Brammer}, {van Dokkum},
  {Illingworth}, {Bouwens}, {Labb{\'e}}, {Franx}, {Momcheva}, {Ashby}, {Fazio},
  {Gonzalez}, {Holden}, {Magee}, {Skelton}, {Smit}, {Spitler}, {Trenti}, \&
  {Willner}}]{Oesch+16}
{Oesch}, P.~A., {Brammer}, G., {van Dokkum}, P.~G., {et~al.} 2016, \apj, 819,
  129, \dodoi{10.3847/0004-637X/819/2/129}

\bibitem[{{Ono} {et~al.}(2023){Ono}, {Harikane}, {Ouchi}, {Yajima}, {Abe},
  {Isobe}, {Shibuya}, {Wise}, {Zhang}, {Nakajima}, \& {Umeda}}]{Ono+23}
{Ono}, Y., {Harikane}, Y., {Ouchi}, M., {et~al.} 2023, \apj, 951, 72,
  \dodoi{10.3847/1538-4357/acd44a}

\bibitem[{{Parlanti} {et~al.}(2023){Parlanti}, {Carniani}, {Pallottini},
  {Cignoni}, {Cresci}, {Kohandel}, {Mannucci}, \& {Marconi}}]{Parlanti+23}
{Parlanti}, E., {Carniani}, S., {Pallottini}, A., {et~al.} 2023, \aap, 673,
  A153, \dodoi{10.1051/0004-6361/202245603}

\bibitem[{{Peng} {et~al.}(2002){Peng}, {Ho}, {Impey}, \& {Rix}}]{Peng+02}
{Peng}, C.~Y., {Ho}, L.~C., {Impey}, C.~D., \& {Rix}, H.-W. 2002, \aj, 124,
  266, \dodoi{10.1086/340952}

\bibitem[{{Peng} {et~al.}(2010){Peng}, {Ho}, {Impey}, \& {Rix}}]{Peng+10}
---. 2010, \aj, 139, 2097, \dodoi{10.1088/0004-6256/139/6/2097}

\bibitem[{{Perna} {et~al.}(2023){Perna}, {Arribas}, {Marshall}, {D'Eugenio},
  {{\"U}bler}, {Bunker}, {Charlot}, {Carniani}, {Jakobsen}, {Maiolino},
  {Rodr{\'\i}guez Del Pino}, {Willott}, {B{\"o}ker}, {Circosta}, {Cresci},
  {Curti}, {Husemann}, {Kumari}, {Lamperti}, {P{\'e}rez-Gonz{\'a}lez}, \&
  {Scholtz}}]{Perna+23}
{Perna}, M., {Arribas}, S., {Marshall}, M., {et~al.} 2023, \aap, 679, A89,
  \dodoi{10.1051/0004-6361/202346649}

\bibitem[{{Perrin} {et~al.}(2014){Perrin}, {Sivaramakrishnan}, {Lajoie},
  {Elliott}, {Pueyo}, {Ravindranath}, \& {Albert}}]{WebbPSF}
{Perrin}, M.~D., {Sivaramakrishnan}, A., {Lajoie}, C.-P., {et~al.} 2014, in
  Society of Photo-Optical Instrumentation Engineers (SPIE) Conference Series,
  Vol. 9143, Space Telescopes and Instrumentation 2014: Optical, Infrared, and
  Millimeter Wave, ed. J.~{Oschmann}, Jacobus~M., M.~{Clampin}, G.~G. {Fazio},
  \& H.~A. {MacEwen}, 91433X, \dodoi{10.1117/12.2056689}

\bibitem[{{Pillepich} {et~al.}(2019){Pillepich}, {Nelson}, {Springel},
  {Pakmor}, {Torrey}, {Weinberger}, {Vogelsberger}, {Marinacci}, {Genel}, {van
  der Wel}, \& {Hernquist}}]{Pillepich+19}
{Pillepich}, A., {Nelson}, D., {Springel}, V., {et~al.} 2019, \mnras, 490,
  3196, \dodoi{10.1093/mnras/stz2338}

\bibitem[{{Planck Collaboration} {et~al.}(2020){Planck Collaboration},
  {Aghanim}, {Akrami}, {Ashdown}, {Aumont}, {Baccigalupi}, {Ballardini},
  {Banday}, {Barreiro}, {Bartolo}, {Basak}, {Battye}, {Benabed}, {Bernard},
  {Bersanelli}, {Bielewicz}, {Bock}, {Bond}, {Borrill}, {Bouchet}, {Boulanger},
  {Bucher}, {Burigana}, {Butler}, {Calabrese}, {Cardoso}, {Carron},
  {Challinor}, {Chiang}, {Chluba}, {Colombo}, {Combet}, {Contreras}, {Crill},
  {Cuttaia}, {de Bernardis}, {de Zotti}, {Delabrouille}, {Delouis}, {Di
  Valentino}, {Diego}, {Dor{\'e}}, {Douspis}, {Ducout}, {Dupac}, {Dusini},
  {Efstathiou}, {Elsner}, {En{\ss}lin}, {Eriksen}, {Fantaye}, {Farhang},
  {Fergusson}, {Fernandez-Cobos}, {Finelli}, {Forastieri}, {Frailis},
  {Fraisse}, {Franceschi}, {Frolov}, {Galeotta}, {Galli}, {Ganga},
  {G{\'e}nova-Santos}, {Gerbino}, {Ghosh}, {Gonz{\'a}lez-Nuevo}, {G{\'o}rski},
  {Gratton}, {Gruppuso}, {Gudmundsson}, {Hamann}, {Handley}, {Hansen},
  {Herranz}, {Hildebrandt}, {Hivon}, {Huang}, {Jaffe}, {Jones}, {Karakci},
  {Keih{\"a}nen}, {Keskitalo}, {Kiiveri}, {Kim}, {Kisner}, {Knox},
  {Krachmalnicoff}, {Kunz}, {Kurki-Suonio}, {Lagache}, {Lamarre}, {Lasenby},
  {Lattanzi}, {Lawrence}, {Le Jeune}, {Lemos}, {Lesgourgues}, {Levrier},
  {Lewis}, {Liguori}, {Lilje}, {Lilley}, {Lindholm}, {L{\'o}pez-Caniego},
  {Lubin}, {Ma}, {Mac{\'\i}as-P{\'e}rez}, {Maggio}, {Maino}, {Mandolesi},
  {Mangilli}, {Marcos-Caballero}, {Maris}, {Martin}, {Martinelli},
  {Mart{\'\i}nez-Gonz{\'a}lez}, {Matarrese}, {Mauri}, {McEwen}, {Meinhold},
  {Melchiorri}, {Mennella}, {Migliaccio}, {Millea}, {Mitra},
  {Miville-Desch{\^e}nes}, {Molinari}, {Montier}, {Morgante}, {Moss}, {Natoli},
  {N{\o}rgaard-Nielsen}, {Pagano}, {Paoletti}, {Partridge}, {Patanchon},
  {Peiris}, {Perrotta}, {Pettorino}, {Piacentini}, {Polastri}, {Polenta},
  {Puget}, {Rachen}, {Reinecke}, {Remazeilles}, {Renzi}, {Rocha}, {Rosset},
  {Roudier}, {Rubi{\~n}o-Mart{\'\i}n}, {Ruiz-Granados}, {Salvati}, {Sandri},
  {Savelainen}, {Scott}, {Shellard}, {Sirignano}, {Sirri}, {Spencer},
  {Sunyaev}, {Suur-Uski}, {Tauber}, {Tavagnacco}, {Tenti}, {Toffolatti},
  {Tomasi}, {Trombetti}, {Valenziano}, {Valiviita}, {Van Tent}, {Vibert},
  {Vielva}, {Villa}, {Vittorio}, {Wandelt}, {Wehus}, {White}, {White},
  {Zacchei}, \& {Zonca}}]{Planck18}
{Planck Collaboration}, {Aghanim}, N., {Akrami}, Y., {et~al.} 2020, \aap, 641,
  A6, \dodoi{10.1051/0004-6361/201833910}

\bibitem[{{Rauscher}(2024)}]{nsclean}
{Rauscher}, B.~J. 2024, \pasp, 136, 015001, \dodoi{10.1088/1538-3873/ad1b36}

\bibitem[{{Rizzo} {et~al.}(2022){Rizzo}, {Kohandel}, {Pallottini}, {Zanella},
  {Ferrara}, {Vallini}, \& {Toft}}]{Rizzo+22}
{Rizzo}, F., {Kohandel}, M., {Pallottini}, A., {et~al.} 2022, \aap, 667, A5,
  \dodoi{10.1051/0004-6361/202243582}

\bibitem[{{Rizzo} {et~al.}(2021){Rizzo}, {Vegetti}, {Fraternali}, {Stacey}, \&
  {Powell}}]{Rizzo+21}
{Rizzo}, F., {Vegetti}, S., {Fraternali}, F., {Stacey}, H.~R., \& {Powell}, D.
  2021, \mnras, 507, 3952, \dodoi{10.1093/mnras/stab2295}

\bibitem[{{Rizzo} {et~al.}(2020){Rizzo}, {Vegetti}, {Powell}, {Fraternali},
  {McKean}, {Stacey}, \& {White}}]{Rizzo+20}
{Rizzo}, F., {Vegetti}, S., {Powell}, D., {et~al.} 2020, \nat, 584, 201,
  \dodoi{10.1038/s41586-020-2572-6}

\bibitem[{{Romeo} \& {Agertz}(2014)}]{Romeo+14}
{Romeo}, A.~B., \& {Agertz}, O. 2014, \mnras, 442, 1230,
  \dodoi{10.1093/mnras/stu954}

\bibitem[{{Romeo} {et~al.}(2010){Romeo}, {Burkert}, \& {Agertz}}]{Romeo+10}
{Romeo}, A.~B., {Burkert}, A., \& {Agertz}, O. 2010, \mnras, 407, 1223,
  \dodoi{10.1111/j.1365-2966.2010.16975.x}

\bibitem[{Romeo \& Falstad(2013)}]{Romeo+13}
Romeo, A.~B., \& Falstad, N. 2013, Monthly Notices of the Royal Astronomical
  Society, 433, 1389–1397, \dodoi{10.1093/mnras/stt809}

\bibitem[{{Romeo} \& {Wiegert}(2011)}]{Romeo+11}
{Romeo}, A.~B., \& {Wiegert}, J. 2011, \mnras, 416, 1191,
  \dodoi{10.1111/j.1365-2966.2011.19120.x}

\bibitem[{{Scholtz} {et~al.}(2023){Scholtz}, {Witten}, {Laporte}, {Ubler},
  {Perna}, {Maiolino}, {Arribas}, {Baker}, {Bennett}, {D'Eugenio}, {Tacchella},
  {Witstok}, {Bunker}, {Carniani}, {Charlot}, {Cresci}, {Curtis-Lake},
  {Eisenstein}, {Kumari}, {Robertson}, {Rodriguez Del Pino}, {Simmonds},
  {Smit}, {Venturi}, {Williams}, \& {Willmer}}]{Scholtz+23}
{Scholtz}, J., {Witten}, C., {Laporte}, N., {et~al.} 2023, arXiv e-prints,
  arXiv:2306.09142, \dodoi{10.48550/arXiv.2306.09142}

\bibitem[{{Tacchella} {et~al.}(2023){Tacchella}, {Eisenstein}, {Hainline},
  {Johnson}, {Baker}, {Helton}, {Robertson}, {Suess}, {Chen}, {Nelson},
  {Pusk{\'a}s}, {Sun}, {Alberts}, {Egami}, {Hausen}, {Rieke}, {Rieke},
  {Shivaei}, {Williams}, {Willmer}, {Bunker}, {Cameron}, {Carniani}, {Charlot},
  {Curti}, {Curtis-Lake}, {Looser}, {Maiolino}, {Maseda}, {Rawle}, {Rix},
  {Smit}, {{\"U}bler}, {Willott}, {Witstok}, {Baum}, {Bhatawdekar}, {Boyett},
  {Danhaive}, {de Graaff}, {Endsley}, {Ji}, {Lyu}, {Sandles}, {Saxena},
  {Scholtz}, {Topping}, \& {Whitler}}]{Tacchella+23}
{Tacchella}, S., {Eisenstein}, D.~J., {Hainline}, K., {et~al.} 2023, \apj, 952,
  74, \dodoi{10.3847/1538-4357/acdbc6}

\bibitem[{{Tokuoka} {et~al.}(2022){Tokuoka}, {Inoue}, {Hashimoto}, {Ellis},
  {Laporte}, {Sugahara}, {Matsuo}, {Tamura}, {Fudamoto}, {Moriwaki},
  {Roberts-Borsani}, {Shimizu}, {Yamanaka}, {Yoshida}, {Zackrisson}, \&
  {Zheng}}]{Tokuoka+22}
{Tokuoka}, T., {Inoue}, A.~K., {Hashimoto}, T., {et~al.} 2022, \apjl, 933, L19,
  \dodoi{10.3847/2041-8213/ac7447}

\bibitem[{Tsukui \& Iguchi(2021)}]{Tsukui+21}
Tsukui, T., \& Iguchi, S. 2021, Science, 372, 1201–1205,
  \dodoi{10.1126/science.abe9680}

\bibitem[{{Walter} {et~al.}(2008){Walter}, {Brinks}, {de Blok}, {Bigiel},
  {Kennicutt}, {Thornley}, \& {Leroy}}]{Walter+08}
{Walter}, F., {Brinks}, E., {de Blok}, W.~J.~G., {et~al.} 2008, \aj, 136, 2563,
  \dodoi{10.1088/0004-6256/136/6/2563}

\bibitem[{{Wisnioski} {et~al.}(2015){Wisnioski}, {F{\"o}rster Schreiber},
  {Wuyts}, {Wuyts}, {Bandara}, {Wilman}, {Genzel}, {Bender}, {Davies},
  {Fossati}, {Lang}, {Mendel}, {Beifiori}, {Brammer}, {Chan}, {Fabricius},
  {Fudamoto}, {Kulkarni}, {Kurk}, {Lutz}, {Nelson}, {Momcheva}, {Rosario},
  {Saglia}, {Seitz}, {Tacconi}, \& {van Dokkum}}]{Wisnioski+15}
{Wisnioski}, E., {F{\"o}rster Schreiber}, N.~M., {Wuyts}, S., {et~al.} 2015,
  \apj, 799, 209, \dodoi{10.1088/0004-637X/799/2/209}

\bibitem[{{Xu} {et~al.}(2024){Xu}, {Ouchi}, {Isobe}, {Nakajima}, {Ozaki},
  {Bouch{\'e}}, {Wise}, {Emsellem}, {Kusakabe}, {Hattori}, {Nagao}, {Chiaki},
  {Fukushima}, {Harikane}, {Hayashi}, {Hirai}, {Kim}, {Maseda}, {Nagamine},
  {Shibuya}, {Sugahara}, {Yajima}, {Aoyama}, {Fujimoto}, {Fukushima}, {Hatano},
  {Inoue}, {Ishigaki}, {Kawasaki}, {Kojima}, {Komiyama}, {Koyama}, {Koyama},
  {Lee}, {Matsumoto}, {Mawatari}, {Moriya}, {Motohara}, {Murai}, {Nishigaki},
  {Onodera}, {Ono}, {Rauch}, {Saito}, {Sasaki}, {Suzuki}, {Takeuchi}, {Umeda},
  {Umemura}, {Watanabe}, {Yabe}, \& {Zhang}}]{Xu+24}
{Xu}, Y., {Ouchi}, M., {Isobe}, Y., {et~al.} 2024, \apj, 961, 49,
  \dodoi{10.3847/1538-4357/ad06ab}

\bibitem[{{Yajima} {et~al.}(2023){Yajima}, {Abe}, {Fukushima}, {Ono},
  {Harikane}, {Ouchi}, {Hashimoto}, \& {Khochfar}}]{Yajima+23}
{Yajima}, H., {Abe}, M., {Fukushima}, H., {et~al.} 2023, \mnras, 525, 4832,
  \dodoi{10.1093/mnras/stad2497}

\bibitem[{{Yajima} {et~al.}(2017){Yajima}, {Nagamine}, {Zhu}, {Khochfar}, \&
  {Dalla Vecchia}}]{Yajima+17}
{Yajima}, H., {Nagamine}, K., {Zhu}, Q., {Khochfar}, S., \& {Dalla Vecchia}, C.
  2017, \apj, 846, 30, \dodoi{10.3847/1538-4357/aa82b5}

\bibitem[{{Yajima} {et~al.}(2022){Yajima}, {Abe}, {Khochfar}, {Nagamine},
  {Inoue}, {Kodama}, {Arata}, {Dalla Vecchia}, {Fukushima}, {Hashimoto},
  {Kashikawa}, {Kubo}, {Li}, {Matsuda}, {Mawatari}, {Ouchi}, \&
  {Umehata}}]{Yajima+22}
{Yajima}, H., {Abe}, M., {Khochfar}, S., {et~al.} 2022, \mnras, 509, 4037,
  \dodoi{10.1093/mnras/stab3092}

\bibitem[{Zhuang \& Shen(2024)}]{Zhuang+24}
Zhuang, M.-Y., \& Shen, Y. 2024, The Astrophysical Journal, 962, 139,
  \dodoi{10.3847/1538-4357/ad1183}

\end{thebibliography}
\bibliographystyle{aasjournal}

\end{document}